\documentclass{aa}  
\usepackage{hyperref}
\usepackage{adjustbox}
\usepackage{arydshln}

\hypersetup{
    colorlinks=true,
    linkcolor=blue,
    filecolor=blue,      
    urlcolor=blue,
    citecolor=blue
}

\usepackage{graphicx}
\usepackage{txfonts}
\newcommand{\GG}[1]{}
\defcitealias{Bh+19}{Paper I}
\defcitealias{Bh+19b}{Paper II}
\defcitealias{Bh+21}{Paper III}
\defcitealias{Bhattacharya22}{Paper IV}

\begin{document}

   \title{The survey of planetary nebulae in Andromeda (M 31) V. Chemical enrichment of the thin and thicker discs of Andromeda} \subtitle{Oxygen to argon abundance ratios for planetary nebulae and H{\rm II} regions}
    \titlerunning{O/Ar abundances in the M~31 thin and thicker discs from PNe}
    \author{Magda Arnaboldi\inst{1} \and 
            Souradeep Bhattacharya\inst{2}\and
          Ortwin Gerhard\inst{3} \and
          Chiaki Kobayashi\inst{4} \and
          Kenneth C. Freeman\inst{5} \and
          Nelson Caldwell\inst{6} \and
          Johanna Hartke\inst{7,8} \and
          Alan McConnachie\inst{9} \and
            Puragra Guhathakurta\inst{10}
          }
   \institute{European Southern Observatory, Karl-Schwarzschild-Str. 2, 85748 Garching, Germany \\ 
   \email{marnabol@eso.org} \and
   Inter University Centre for Astronomy and Astrophysics, Ganeshkhind, Post Bag 4, Pune 411007, India \and
   Max-Planck-Institut für extraterrestrische Physik, Giessenbachstraße, 85748 Garching, Germany \and
   Centre for Astrophysics Research, Department of Physics, Astronomy and Mathematics, University of Hertfordshire, Hatfield, AL10 9AB, UK \and
      Research School of Astronomy and Astrophysics, Mount Stromlo Observatory, Cotter Road, ACT 2611 Weston Creek, Australia \and
      Harvard-Smithsonian Center for Astrophysics, 60 Garden Street, Cambridge, MA 02138, USA \and
   European Southern Observatory, Alonso de C\'ordova 3107, Santiago de Chile, Chile \and   
   Sub-Department of Astrophysics, Department of Physics, University of Oxford, Denys Wilkinson Building, Keble Road, Oxford OX1 3RH, UK \and
   NRC Herzberg Institute of Astrophysics, 5071 West Saanich Road, Victoria, BC V9E 2E7, Canada \and
   UCO/Lick Observatory, Department of Astronomy \& Astrophysics, University of California Santa Cruz, 1156 High Street, Santa Cruz, California 95064, USA
             }

   \date{Accepted 16.07.2022. Received 13.06.2022; in original form 13.06.2022}



\abstract

\abstract
{The Andromeda (M~31) galaxy presents evidence of a recent substantial mass accretion, differently from the Milky Way that had a rather quiescent evolution.} 
{We use oxygen and argon abundances for planetary nebulae (PNe) with low internal extinction (progenitor ages of $>4.5$ Gyr) and high extinction (progenitor ages $<2.5$ Gyr), as well as those of the H{\rm II} regions, to constrain the chemical enrichment and star formation efficiency in the thin and thicker discs of M31. }
{The argon element is produced in larger fraction by Type Ia supernovae (SNe) than oxygen. We find that the mean log(O/Ar) values of PNe as a function of their argon abundances 12 + log(Ar/H) trace the inter-stellar matter (ISM) conditions at the time of birth of the M~31 disc PN progenitors. Thus the chemical enrichment and star formation efficiency information encoded in the [$\alpha$/Fe] vs. [Fe/H] distribution of stars is also imprinted in the oxygen-to-argon abundance ratio log(O/Ar) vs. argon abundance for the nebular emissions of the different stellar evolution phases. We propose to use the log(O/Ar) vs. (12 + log(Ar/H)) distribution of PNe with different ages to constrain the star-formation histories of the parent stellar populations in the thin and thicker M31 discs. } 
{For the inner M31 disc ($R_{GC} < 14$ kpc), the chemical evolution model that reproduces the mean log(O/Ar) values as function of argon abundance for the high- and low-extinction PNe requires a second infall of metal poorer gas during a gas-rich (wet) satellite merger. This wet merger triggered the burst of star-formation seen by the PHAT survey in the M31 disc, $\sim 3$ Gyr ago. {  A strong starburst is on going in the intermediate radial range ($14 \le R_{GC} \le 18$ kpc )}. In the outer M31 disc ($R_{GC} >18$ kpc), the log(O/Ar) vs argon abundance distribution of the younger high extinction PNe indicates that they too were formed in a burst, but mostly from the metal poorer gas. Present-day H{\rm II} regions show a range of oxygen to argon ratios indicative of spatial variations, consistent with a present day rainfall of metal poorer gas onto the disc with different degree of mixing with the previously enriched ISM.}
{We implement the use of the log(O/Ar) vs argon abundance distribution for emission nebulae as a complement to the [$\alpha$/Fe] vs. [Fe/H] diagram for stars, and use it to constrain the star formation efficiency in M31 thin and thicker discs. Diagrams for M31 PNe in different age ranges reveal that a secondary infall of gas affected the chemical evolution of the M31 disc. In M31, the thin disc is {  younger and less radially extended}, formed stars at a higher star formation efficiency, and had a faster chemical enrichment timescale than the {  more extended, thicker} disc.  Both the thin and thicker disc in M31 reach similar high argon abundances $( 12 + \log(Ar/H) )\simeq 6.7$. {  The chemical and structural properties of the thin/thicker discs in M31 are thus remarkably different from those determined for the Milky Way thin and thick discs.} } 


   \maketitle
%

\section{Introduction}
\label{sect:intro}

Late-type galaxies may contain kinematically distinct components such as the ``cold" thin disc and the ``hot" thick disc found in the Milky Way \citep[MW; e.g.][]{Gilmore83} and in nearby galaxies \citep{Yoachim06, Comeron19}. The MW thick disc is {  structurally} and chemically distinct from the MW thin disc {  as well as in age.} {  Differences in the structural parameters (stellar masses, exponential scale lengths, and velocity dispersion) of the MW thin and thick discs are summarized in \citet{bhg16}.}  The different chemical properties are most prominent in the measured stellar [$\alpha$/Fe] ratios as a function of [Fe/H], with the old MW thick disc being more metal-poor and $\alpha$-enriched, compared to the relatively younger MW thin disc \citep{Hayden15,Matteucci21}. {The $\alpha$-enriched thick disk population is found to be confined within $R\simeq 9$ kpc, and older than $\sim 8$ Gyr \citep{Haywood13,Belokurov20}, whereas the outer MW disc is composed of low [$\alpha$/Fe] stars \citep{Hayden15}.} These properties are believed to have been set by the MW's most recent impactful merger $\sim$10 Gyr ago \citep{Belokurov18,Helmi18}, after which the MW disc is thought to have evolved mainly by secular evolution \citep[see, e.g.,][]{Sellwood14}.

Differently from the MW, M~31 had a more turbulent history, as vividly illustrated by the many substructures identified in its inner halo by PAndAS \citep{mcc09,mcc18}, including the Giant Stellar Stream (GSS, \citealt{Ibata01}). Its most recent important merger is believed to have happened $\sim$ 2.5--4.5 Gyr ago (\citealt[][]{Bh+19b}, hereafter Paper II).
{The M31 disc has a significantly steeper age-velocity dispersion (AVD) relation than that of the MW disc (\citetalias{Bh+19b}; \citealt{dorman15}), with the velocity dispersion of the 2.5 Gyr and 4.5 Gyr old stellar populations being almost twice resp.\ three times those of the MW disc stellar populations of corresponding ages.} \citetalias{Bh+19b} used planetary nebulae as kinematic tracers to identify a younger (progenitor ages $<2.5$ Gyr), dynamically colder disk, and a distinct, older (progenitor ages $>4.5$ Gyr), dynamically hotter, hence thicker disc, with the latter having a velocity dispersion $\rm\sigma_{M31,thick} \simeq  3 \times \sigma_{MW,thick}$ in the radial range 14-20 kpc (equivalent to the solar neighbourhood). At these  radial distances (R$\rm_{GC}=$14--20 kpc), the 4.5 Gyr and older population of stars had velocity dispersion values $ \geq 90$ kms$^{-1}$, which are significantly larger even than the average velocity dispersion measured for strongly turbulent discs at redshift $\sim1-2$, 30 and 60 kms$^{-1}$, respectively, see \citet{Wisnioski15}.
Mergers with satellites can dynamically heat thin discs, i.e., increase their velocity dispersion \citep{Quinn86} and decrease their rotational velocity, resulting in a thickened disc \citep{Hopkins09}. Using their results, the AVD relation of the M~31 disc in a radial range R$\rm_{GC}=$14--20 kpc was found to be consistent with the energy injected in the M~31 disc by a major merger with mass ratio $\sim$1:5 $\sim$2.5--4.5 Gyr ago \citep{Bh+19b}, as predicted in the merger simulations of \citet{ham18}.

The different merger histories of these two spiral galaxies are seen in their large separation  in the halo metallicity vs.\ total stellar mass diagram, where the MW and M31 are placed at opposite edges of the distribution of galaxies measured by the GHOST survey \citep{Monachesi2019}. In simulations, the spread in halo masses and [Fe/H] values is found to be indicative of different accretion histories \citep{DSouza18b} and accreted satellite stellar mass. Independent confirmation of a recent major merger event in M31 can be sought through chemical abundances, altered by merger-related processes such as gas accretion and star formation bursts \citep[e.g.][]{Kobayashi11}.

Planetary nebulae (PNe) are useful tracers to constrain the kinematics \citep{Aniyan18, Aniyan21} and chemical abundances \citep{Magrini16,Stanghellini18} over a large radial range in nearby galaxies of different morphological types \citep[e.g.][]{cor13,pul18,Hartke22}. PN elemental abundances shed light on the ISM conditions at the time of formation of their parent stellar population. When the PN ages are also constrained, it becomes possible to map abundance variations across different epochs of star formation in galaxies. Abundance distributions and gradients in galaxies were measured using PNe \citep{Maciel94, Magrini16, Kwitter21}. In the MW, {  negative} radial oxygen abundance gradient for both thin and thick disc were constrained using PNe \citep{Stanghellini18}, {  indicating inside-out disk formation. With this aim a large survey of M31 planetary nebulae was undertaken \citep{Bh+19}. While it is not possible to determine the} [$\alpha/$Fe] {  abundance ratio in PNe for constraining the chemical evolution,}  different chemical enrichment timescales which result from distinct star-formation histories also leave imprints in the abundance distribution of other elements \citep[see][and references therein]{Nomoto13}.

{\citet[][hereafter Paper IV]{Bhattacharya22} measured distinct oxygen and argon abundance distributions for the thin and thicker discs in M31. They measured a flat or slightly positive oxygen and argon gradients for the older, thicker disc, and a negative metallicity gradient for the younger thin disc in M31.} These results are consistent with a major merger with 1:5 mass ratio for M31.  In the simulations of \citet{ham18}, a gas-rich satellite is accreted on to M~31 with an orbit along the GSS. The merger then heats the pre-existing M31 disc, generating the observed thicker disc. The cold gas accreted through the ``wet'' merger would lead to a burst of star formation and the formation of a late, more centrally concentrated thin disc.  Because of the merger--driven disc evolution, stars in the thin and thicker discs of M~31 would have formed at different epochs under different chemical conditions. Those in the younger thin disc would have formed out of the pre-enriched interstellar gas in M31, mixed with metal-poorer gas brought in by the satellite. 

In this paper we investigate whether the different chemical enrichment of the thicker and thin discs in M31 provide ``smoking gun" evidence that the secondary gas infall predicted by such a gas-rich merger took place.  We will use the direct measurements of oxygen and argon abundances for the M~31 disc PNe over the $2- 30$ kpc radial range to constrain the chemical enrichment and the star formation efficiency in the thin and thicker disc. We further combine the PNe abundance measurements with those for the H{\rm II} regions in the M31 disc already available in the literature. The oxygen and argon measurements from \citetalias{Bhattacharya22} are briefly presented in Section~\ref{sect:data}. The use of oxygen and argon abundances as chemical tracers in M~31 is discussed in Section~\ref{sect:arg_dep}, as well as the chemical enrichment timescales inferred for the two disc components. In Section~\ref{sect:disc2} we present the constraints on the chemical evolution and formation history of M~31. We address the gas content of the merging satellite in the M~31 disc in Section~\ref{sect:satgas}, and conclude in Section~\ref{sect:future}.

\section{Data sample, abundance measurements and gradients}
\label{sect:data}

\subsection{The PN sample in M31 with O and Ar abundances }
\label{sect:obs}

In \citet{Bh+19}, PN candidates were selected in a 16 sq. deg. [OIII] 5007 \AA\ narrow plus g-broad band imaging survey covering the disc and inner halo of M~31, with MegaCam at the CFHT. This was later expanded to cover a total of 54 sq. deg in M31 \citep{Bh21}.  Spectroscopic follow-up of a complete sub-sample of these PN candidates were carried out with the Hectospec multifibre positioner and spectrograph on the Multiple Mirror Telescope \citep[MMT;][]{fab05}. Spectral range covers from 3685\AA\  in the blue to approx 9200 \AA\ in the red, with a spectral resolution in the range $850-1500$. 

The oxygen and argon direct abundance measurements, via the detection of the temperature sensitive line  [\ion{O}{iii}] at $4363$ ~\AA~, with their errors, for a magnitude limited sample of PNe in the M31 disc, are described in \citetalias{Bhattacharya22} which includes the catalogue with the measured quantities (their errors). 
We also refer to \citetalias{Bhattacharya22} for a comprehensive description of the PNe magnitude limited sample and the abundance gradients. In this work we extend the analysis of the oxygen and argon abundances to constrain the chemical enrichment and the star formation timescales of the discs of M31.

{To this aim, we implement the identification of the thin/thicker discs in M31 of \citet{Bh+19b}, based on the ages and dynamical properties of their PN populations, and then derive their star formation histories and chemical evolution. In the M31 magnitude-limited disc PNe sample, there are 75 high-extinction, 2.5 Gyr and younger PNe associated with the more rapidly rotating thin disc, and 130 low extinction, 4.5 Gyrs and older PNe,  associated with the thicker disc, which has a larger asymmetric drift.  
The high extinction PNe are found at smaller radii ($R_{GC} < 22$ kpc) than the low extinction PNe, which instead cover the entire extent of the disc (out to $R_{GC}=30$ kpc).}

{We note that the structural properties of thin/thick discs in nearby spiral galaxiess show thin discs embedded in thicker discs, with the latter having longer scale lengths than the thin component \citep{Yoachim06}. This is similar to what is found for the M31 PNe 2.5 Gyr and younger sample. Furthermore the age determined for the thicker disk in M31 falls within the age range, from 4 to 10 Gyr, determined from Lick indices for the thick discs of local spirals \citep{Yoachim08}.}

\subsection{The PN sample in M31 as tracers of the ISM chemical properties}
\label{sect:agbism}

As a first step in using the PN oxygen over argon ratio vs. argon abundance, 12 + log(Ar/H), distribution to constrain the ISM conditions, we explore the dependency of the PNe log(O/Ar) ratio on i) their circumstellar dust properties, ii) initial stellar masses and iii) ages. 
 According to \citet{Ventura17} theoretical models, the surface oxygen in the stellar atmospheres maybe modified during the AGB stage in two ways:
 \begin{itemize}
     \item oxygen in the stellar atmosphere can be enriched by the Third Dredge Up (TDU) which may occur in 2 Gyr old stars, with masses between $ 1.5-3 M_\sun$, and metallicity range $[-0.5: 0.0]$.  In the case of the M31 disc low extinction PNe, which are 4.5 Gyr and older, with low mass $M_* < 1.25 M_\sun$ (see \citetalias{Bh+19b}), their measured oxygen abundance are free of alterations from PAGB evolution, and TDU in particular.
 \item oxygen can be depleted by hot bottom burning (HBB) which occurs in $M_*> 3.5 M_\sun$ progenitors during the AGB phase. In \citetalias{Bhattacharya22} (see their Appendix D), we found that the contribution to the M31 disc PN sample from very young (<300 Myr old) massive ($M> 3.5 M_\sun$) stars is negligible, therefore there is no evidence for oxygen depletion in the current M31 disc high extinction PN sample. 
 \end{itemize}
In summary, the current sample of low extinction PNe in the M31 disc are too old and of low masses for TDU to occur, while there is no evidence for the presence of very young (300 Myr) massive ($> 3.5 M_\odot$) PNe in the current high extinction PNe sample in M31, which may be affected by oxygen depletion. 
We refer to Appendix~\ref{sect:agb} for more in depth discussion of oxygen and argon abundances in PNe. 
  
Differently from oxygen, argon is known to be invariant during the AGB evolution \citep{Delgado14,Garcia-Hernandez16,Ventura17}.  Given the absence of oxygen modifications due to PAGB effects in the current M31 disc PNe sample, in what follow we then proceed to use the PN oxygen and argon abundances to study the chemical properties of the ISM at the time the PN stellar progeny were formed.

\subsection{Oxygen and argon radial gradients for thin and thicker disc in M31}
\label{sect:grad}

In \citetalias{Bhattacharya22}, we measured two distinct abundance distributions for the oxygen and argon, for the thin and thicker disc in M31. The mean value of the oxygen abundance for the thin disc, $\rm <12+(O/H)>_{high-ext}= 8.57 \pm 0.03$, is higher than that of the thicker disc, $\rm <12+(O/H)>_{low-ext.} = 8.48 \pm 0.02$, although both distributions have large standard deviation values. Same trend for the mean values of the argon abundance: the argon abundance of the thin disc, $\rm <12+(Ar/H)>_{high-ext} = 6.32 \pm 0.03$, is higher than that of the thicker disc, $\rm <12+(O/H)>_{low-ext} = 6.25 \pm 0.02$. When the two abundance distributions are compared in pairs, the two-sample Anderson-Darling test rejects the null hypothesis that the two distributions of each element are drawn from the same, underlying, distribution. Hence the two discs in M31 are chemically distinct in oxygen abundance distribution and argon abundance distribution. 

Regarding the abundance gradient with radius, in \citetalias{Bhattacharya22} we find a steeper negative radial  gradient for the oxygen abundance for the thin disc, $\rm (\Delta(O/H)/\Delta R)_{high-ext}= -0.013\pm0.006$ dex/kpc, which is consistent with that measured for the HII regions \citep{Zurita12}. \citetalias{Bhattacharya22} also measured a near-flat and slightly positive radial gradient for the oxygen abundance of the thicker disc, $\rm (\Delta(O/H)/\Delta R)_{low-ext} = 0.06\pm0.003$ dex/kpc. The measured radial gradients for the argon abundance are $\rm (\Delta(Ar/H)/\Delta R)_{high-ext}= -0.018\pm0.006$ dex/kpc and $\rm (\Delta(O/H)/\Delta R)_{high-ext}= -0.05\pm0.003$ dex/kpc respectively.  The results of \citetalias{Bhattacharya22} are consistent with results of previous studies \citep{san12,Kw12,Pena19}, whose oxygen gradient measurements were dominated by the more numerous low-extinction PNe associated with the thicker disc. We refer to \citetalias{Bhattacharya22} for a more extensive comparison of these radial gradients with those of the MW and other spirals.

\begin{figure}
        \centering
        \includegraphics[width=\columnwidth,angle=0]{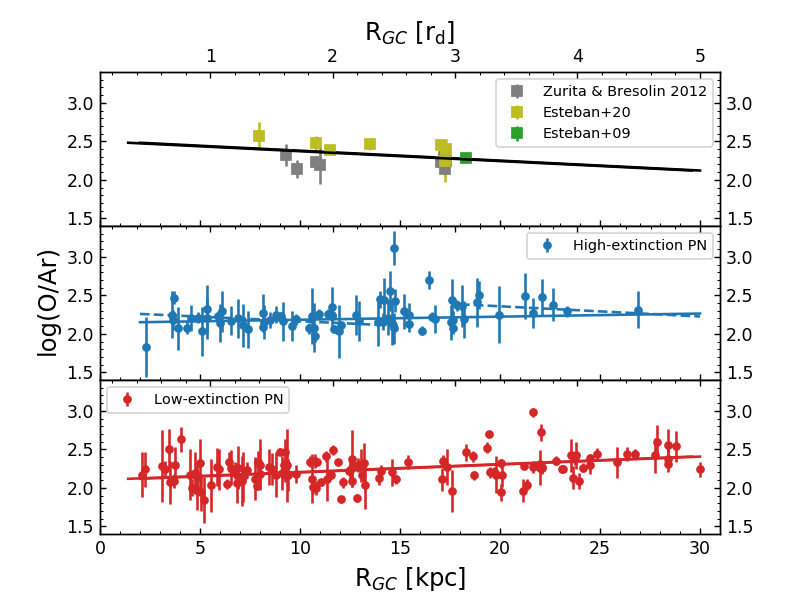}
        \caption{The galactocentric radial distribution of log(O/Ar) values for [top] HII regions, [middle] high- and [bottom] low-extinction PNe in the M~31 disc from the PN catalogue published in \citetalias{Bhattacharya22}. The best-fitting radial log(O/Ar) gradient is shown for HII regions  (black), high- (blue) and low-extinction (red) PNe.  The middle panel displays three independent linear fits for three radial ranges: the solid line is the linear fit to the entire data; for within 14 kpc and beyond 18 kpc, the linear fits are shown with dashed lines.
        }
        \label{fig:oar_grad}
\end{figure}

\begin{table}
\caption{Fitted parameters for the radial gradients of the log (O/Ar) values for HII regions (literature values) and in the M~31 disc from the \citetalias{Bhattacharya22} M~31 disc PNe sample. }
\centering
\adjustbox{max width=\columnwidth}{
\begin{tabular}{ccccc}
\hline
Sample & $\log(O/Ar)\rm_0$ & \multicolumn{2}{c}{$\rm\Delta \log(O/Ar)/\Delta R$}\\
 &  & dex/kpc & dex/r$\rm_{d}$\\
\hline
\\
HII regions  & 2.51 $\pm$ 0.13 & -0.013 $\pm$ 0.008 & -0.078 $\pm$ 0.049  \\
\hdashline
\\
High-extinction PNe (all) & 2.14 $\pm$ 0.04 & 0.004 $\pm$ 0.003 & 0.024 $\pm$ 0.021  \\
 R$\rm_{GC}\leq14$~kpc & 2.28 $\pm$ 0.05 & -0.011 $\pm$ 0.005 & -0.072 $\pm$ 0.032  \\
R$\rm_{GC}>18$~kpc & 2.62 $\pm$ 0.28 & -0.013 $\pm$ 0.012 & -0.08 $\pm$ 0.075  \\
\hdashline
\\
Low-extinction PNe  & 2.1 $\pm$ 0.05 & 0.01 $\pm$ 0.003 & 0.063 $\pm$ 0.016  \\
\hline
\end{tabular}
\label{table : oxyfit}
}
\end{table}

\subsection{Radial gradients of the log(O/Ar) values}
\label{sect:oar_grad}

Oxygen and argon abundances are reliably measured in the M31 PNe (see  Section~\ref{sect:agbism}, Appendix~\ref{sect:agb} and \citetalias{Bhattacharya22}). Since PNe evolve from parent stellar populations covering a range of ages, their log(O/Ar) ratios probe the ISM conditions at the different epochs of their birth, and thus provide important constraints for the chemical evolution models of the ISM (see review by \citealt{Nomoto13} and work by \citealt{Kobayashi20}).

In M31 we identified three population of tracers in three different age ranges. The HII regions are tracing the chemistry of the  $\sim 300$ Myr young stellar population, while the high extinction PNe, which trace the thin disc of M 31, have ages about 2.5 Gyrs or younger. The low-extinction PNe that are associated with the thicker disc of M~31, are 4.5 Gyrs or older; see \citetalias{Bh+19b} and \citetalias{Bhattacharya22} for further details.

Figure~\ref{fig:oar_grad} shows the galactocentric radial distribution of the log(O/Ar) values for H{\rm II} regions\footnote{Their oxygen and argon abundances were also determined directly using the [\ion{O}{iii}] temperature sensitive line at 4363 ~\AA\ (see \citet{Esteban09,Zurita12,Esteban20}) within a R$\rm_{GC}=17$~kpc radius. Because H{\rm II} regions have high internal extinction, direct detection of the [\ion{O}{iii}] 4363 ~\AA~ emission line may lead to select preferentially a relatively metal-poorer population.}  (upper panel),  high-extinction PNe (middle panel) and low-extinction (lower panel) PN samples, in the  R$\rm_{GC}=$~2--30 kpc radial range. Parameters of linear fits are also noted in Table~\ref{table : oxyfit}. We find a negative radial gradient for the H{\rm II} regions, with relatively large error bars. For the low-extinction PNe, we determine a slightly positive radial gradient, with a mixture of high and low log(O/Ar) values at any radius over the $2-20$ kpc radial range of the disc. For the high-extinction PNe, the figure indicates distinct log(O/Ar) distributions in three radial regions. Within $R_{GC}< 14$ kpc, $\rm<\log(O/Ar)> = 2.28\pm 0.05$ with a slightly negative radial gradient. In the $R_{GC}>18$ kpc outer region $\rm <\log(O/Ar)> = 2.62\pm 0.28$ with a negative radial gradient as for the inner region. In the intermediate radial range $14-18$ kpc, which includes the most active star forming region in the M31 disc \citep{Kang09},  the log(O/Ar) values have a wide  spread, including the largest inferred values ($> 2.5$) for the PNe in our sample. 
We discuss the radial gradients and the constraints from the log(O/Ar) distributions of the different age ranges on the chemical evolution models for the thin and thicker disc in M31 in the following sections.

\section{Oxygen vs argon as tracers of the enrichment history in the M~31 discs}
\label{sect:arg_dep}

The star-formation history of a galaxy leaves chemical imprints in its ISM through enrichment with different elements. The information of the ISM chemical conditions at the time of birth of a star is encoded in its element abundances. While both argon and oxygen are produced from core-collapse supernovae, argon is additionally produced by Type-Ia supernovae \citep{Kobayashi20,Kobayashi20b}. Thus even though both argon and oxygen are $\alpha$-elements, they do not have lockstep behavior and the star-formation histories of parent stellar populations will imprint information on the log(O/Ar) vs. argon abundance, 12 + log(Ar/H), distribution of stars with different ages.

\subsection{M~31 PNe distribution of the in the log(O/Ar) vs. 12 log(Ar/H) plane}
\label{sect:trends}

\begin{figure*}
        \centering
        \includegraphics[width=\columnwidth,angle=0]{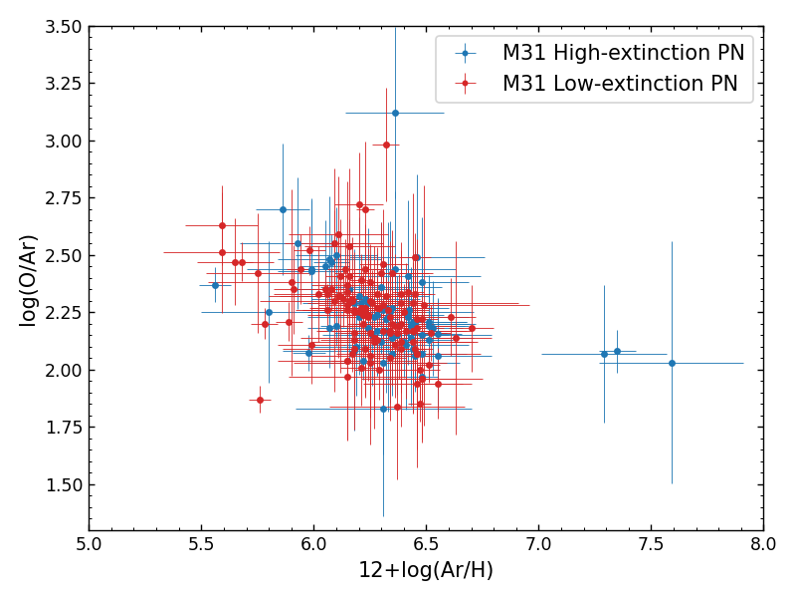}
        \includegraphics[width=\columnwidth,angle=0]{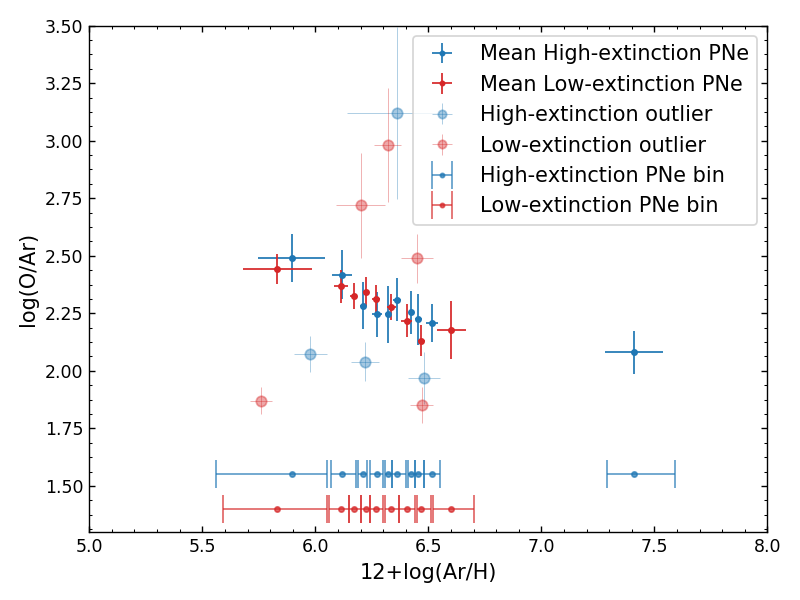}
        \caption{[Left] Oxygen-to-argon abundance ratio, log(Ar/O), plotted against argon abundance for the high- (blue) and low- (red) extinction PNe in M~31. [Right] high- (blue) and low- (red) extinction PNe binned separately as a function of their 12+log(Ar/H) values. For each bin the sigma-clipped mean log(O/Ar) is plotted as well its error (error bar along the y-axis), see text section~\ref{sect:trends}. The error is computed as the error on the mean log(O/Ar) value added in quadrature to the mean observation error of the PNe which remain in each bin, after the sigma-clipping. Clipped PN with log(O/Ar) values that are more than 2-sigma beyond the mean values are marked with a lighter shade. Error bars along the x-axis capture the rms of PNe in each argon abundance bin. Red and blue error-bars parallel to the x-axis indicate the width of the bins {  in 12+log(Ar/H)} for the low-extinction (red) and high extinction (blue) PN samples. }
        \label{fig:arh_oh}
\end{figure*}

\begin{table}
\caption{log(O/Ar) average values in the 12+log(Ar/H) bins, for the low extinction Pne in M31. }
\centering
\adjustbox{max width=\columnwidth}{
\begin{tabular}{ccccc}
\hline
\\
 12+log(Ar/H) &	log(O/Ar) &	$1 \times \sigma $	& Error on  & Mean measurement \\
 &	&	log(O/Ar)	&  mean log(O/Ar) & error on log(O/Ar)\\
\\\hline\\
5.83 &	2.44 &	0.10 &	0.02 &	0.07 \\
6.11 &	2.37 &	0.10 &	0.03 &	0.08 \\
6.17 &	2.33 &	0.09 &	0.02 &	0.06 \\
6.22 &	2.34 &	0.12 &	0.03 &	0.07 \\
6.27 &	2.31 &	0.10 &	0.02 &	0.07 \\
6.34 &	2.28 &	0.12 &	0.03 &	0.06 \\
6.40 &	2.22 &	0.09 &	0.02 &	0.08 \\
6.47 &	2.13 &	0.16 &	0.04 &	0.07 \\
6.60 &	2.18 &	0.10 &	0.03 &	0.12 \\
\\\hline
\end{tabular}
\label{table:LE_errors}
}
\tablefoot{Listed values include standard deviation of log(O/Ar) in each bin, error on mean log(O/Ar) values and mean measurement error in each bin.
}
\end{table}

\begin{table}
\caption{log(O/Ar) average values in the 12+log(Ar/H) bins, for the high extinction PNe in M31.}
\centering
\adjustbox{max width=\columnwidth}{
\begin{tabular}{ccccc}
\hline
\\
 12+log(Ar/H) &	log(O/Ar) &	$1 \times \sigma $	& Error on  & Mean measurement \\
 &	&	log(O/Ar)	&  mean log(O/Ar) & error on log(O/Ar)\\
 \\
\hline\\
5.89 &	2.49 &	0.12 &	0.05 &	0.09\\
6.12 &	2.42 &	0.13 &	0.05 &	0.10\\
6.21 &	2.28 &	0.09 &	0.03 &	0.09\\
6.27 &	2.24 &	0.10 &	0.04 &	0.09\\
6.32 &	2.25 &	0.11 &	0.04 &	0.12\\
6.36 &	2.31 &	0.12 &	0.04 &	0.08\\
6.42 &	2.25 &	0.12 &	0.04 &	0.08\\
6.46 &	2.23 &	0.09 &	0.03 &	0.10\\
6.52 &	2.21 &	0.11 &	0.04 &	0.07\\
7.41 &	2.08 &	0.02 &	0.01 &	0.09\\
\\\hline
\end{tabular}
\label{table:HE_errors}
}
\tablefoot{Listed values include standard deviation of log(O/Ar) in each bin, error on mean log(O/Ar) values and mean measurement error in each bin.
}
\end{table}

The left panel of Figure~\ref{fig:arh_oh} shows the oxygen-to-argon abundance ratio, log(O/Ar), vs. 12+log(Ar/H) for all the M~31 PNe. 
In order to identify the general trend in the log(O/Ar) vs. argon abundance, we adopt the following procedure. We divide the high- and low- extinction PNe in bins of argon abundance, 12+log(Ar/H), such that there are between $8$ and $15$ PNe in each bin. The number of PNe in each bin are chosen such that there are enough PNe in each bin to obtain a reliable dispersion in log(O/Ar) while covering the entire 12+log(Ar/H) range with as many bins as possible. The number of PNe in each bin is lower for the high-extinction PNe as they are fewer in total\footnote{Note that the bin with the largest argon abundance has fewer measurements (three) for the high-extinction PNe.}. We then calculate the mean log(O/Ar) values of PNe in each 12+log(Ar/H) bin after removing the $2 \times \sigma$ outliers from the mean\footnote{In practice, we first calculate the mean and standard deviation from all PNe in each bin, then remove those PNe whose log(O/Ar) values are more than $2 \times \sigma$ from the mean, and recompute the mean value for the clipped sample}. 
In Table~\ref{table:LE_errors} and Table~\ref{table:HE_errors} we provide the mean values of log(O/Ar) in the 12+log(Ar/H) bins, their standard deviation in bins, the error on the mean value and mean measurement error in each bin, for the low and high extinction PNe samples. We discuss them in turn. 

In Table~\ref{table:LE_errors}, we compare the standard deviation with the mean measurement error in each bin for the low extinction PNe: the scatter is of order $\sqrt{2}\times$ the mean measurement error for most bins. We thus infer that there is an intrinsic scatter of the measured log(O/Ar) for the low extinction PN sample, and this scatter is of a similar order of magnitude to the mean error in the bins. We thus adopt as the error for the log(O/Ar) average value, in each bin, the error on the mean log(O/Ar) value added in quadrature to the mean observation error of the PNe, in each bin.

For the high extinction PNe in Table~\ref{table:HE_errors}, we notice that the standard deviation is smaller than the average measured error for about half of the bins. We still adopt as the error for the log(O/Ar) average value, in each bin, the error on the mean log(O/Ar) value added in quadrature to the mean observational error in each bin, keeping in mind that the dispersion of values is smaller for 12+log(Ar/H) range 6.21 - 6.32, and for the larger [Ar/H] abundance values. Finally in the right panel of Figure~\ref{fig:arh_oh}, we show the resulting mean log(O/Ar) values in bins of 12+log(Ar/H) with the adopted error bars. 

The distribution in the right panel of Figure~\ref{fig:arh_oh} is clearly a function of argon abundance, with the highest log(O/Ar) values ($\simeq 2.5$) at the lowest measured argon abundance (12+log(Ar/H)$\simeq 6.0$) and the lowest mean log(O/Ar) measured value ($\simeq 2.08$) for the most argon abundant (12+log(Ar/H)$\simeq 7.5$) bin, for the high extinction PNe. We also show the clipped outlier PNe individually. They are few in number and do not affect the general trend. 

\subsection{Tracing chemical enrichment in the old disc with PN abundances}
\label{sect:olddisc}
 
\begin{figure}
        \centering
        \includegraphics[width=\columnwidth,angle=0]{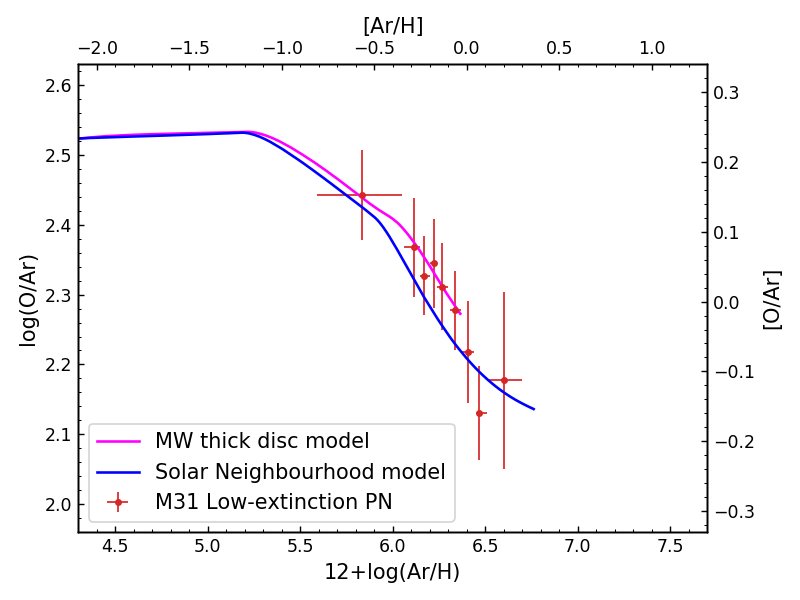}
        \caption{ log(O/Ar) values for the low-extinction PNe in M31 are binned as a function of their 12+log(Ar/H) values and shown as red symbols; for each bin the mean log(O/Ar) is plotted as well as the sigma-clipped error. The continuum lines reproduce the chemical evolution tracks for the oxygen and argon elements modelled by \citet{Kobayashi20}, for the MW thick disc (magenta) and the solar neighbourhood (dark blue). The [Ar/H] abundances and [O/Ar] ratios with respect to the solar values    ($12+\log(Ar/H)_\odot=6.38$ and $\log(O/Ar)_\odot=2.29$). are marked on the top and right axis, respectively.}
        \label{fig:chem}
\end{figure}

 We now focus on the distribution of, and trend in, the mean log(O/Ar) values vs. 12+log(Ar/H) for the low-extinction PNe, which trace the early evolution. In Figure~\ref{fig:chem}, their values are compared with the evolution tracks for oxygen and argon for the MW thick and thin discs \citep{Kobayashi20}. According to these models, argon is made in both SN~II and SN~Ia, with a greater fraction in SN~II than for Fe and a smaller fraction than for oxygen. Therefore the decrease in the log(O/Ar) vs. 12+log(Ar/H) plane would signal the increase in the SN~Ia contribution, which is indeed shown by the models. In the solar neighborhood, 34\% of Ar is produced from SN~Ia by present, see \citet{Kobayashi20}. Interestingly, the \citet{Kobayashi20} model tracks for MW thin and thick discs show two knees: the second knee, at [Ar/H] $= -0.5$, is caused by the contribution of Type Ia SN, similarly to that seen in $\alpha/Fe$ vs Fe/H plots as in \citet{Hayden15}, while the knee at [Ar/H]  $  = -1.2$ is caused by the dependency of argon production on the metallicity of core collapsed Type II SNe.

We first compare the log(O/Ar) vs. 12+log(Ar/H) binned distribution for low-extinction PNe with the chemical evolution models. The O and Ar abundance distributions of the M31 low-extinction disc PNe show a possible slightly positive and null (within the errors) radial gradient, respectively, see \citetalias{Bhattacharya22}. In turn, the log(O/Ar) distribution has a near constant average values and similar large scatter over the entire 2-30 kpc range  (see Section~\ref{sect:oar_grad}, Fig.~\ref{fig:oar_grad} and Table~\ref{table : oxyfit} ). 
{  The measured null radial abundance gradients for the low-extinction PNe and the large and constant scatter across the entire radial range supports the view that their distribution in the log(O/Ar) vs. 12+log(Ar/H) plane is the result of an effective radial mixing of different chemical enrichment tracks over the $2-30$ kpc radial range. Significant radial mixing of the old stellar population over the entire PHAT area was also reported by \citet{wil17}.}

In Figure~\ref{fig:chem}, the comparison of the log(O/Ar) vs. 12+log(Ar/H) binned distribution for low-extinction PNe with the \citet{Kobayashi20} models shows that the highest log(O/Ar) at low 12 + log(Ar/H) values for M~31 disc low-extinction PNe are representative of the stellar population that forms soon after $\alpha$ elements are produced in core collapse SNs, from short-lived massive stars. After Type Ia supernovae start producing additional argon relative to oxygen, the ISM is enriched and stars subsequently form with decreasing log(O/Ar) values at increasing argon abundance (12 + log(Ar/H)). 
The same chemical evolution models, following the aforementioned processes, also show the decreasing trends in the analogous [$\alpha$/Fe] vs. [Fe/H] plot for stellar abundances, as seen for the MW \citep[see e.g.][]{Hayden15}.

In the MW (also recently in UGC~10738; \citealt{Scott21}), it was found that the thick disc is more metal-poor and $\alpha$-enriched compared to the thin disc, occupying distinct regions of the [$\alpha$/Fe] vs. [Fe/H] plot \citep[see e.g.][]{Hayden15,Kobayashi20}. The explanation for the $\alpha$ enhancement in the MW thick disc is that it has had a faster chemical enrichment timescale than the MW thin disc, with high star-formation efficiency in the thick disc at early times, while the extended star formation in the latter generates higher iron abundance values than those in the MW thick disc.
Figure~\ref{fig:chem} shows that correspondingly the MW thin disc also reaches higher Ar abundances than the thick disc.  On the contrary, differently from the MW thick disc, the 4.5 Gyr and older thicker disc in M31 reaches higher argon abundances than the MW thick disc and as high as those reached by the extended star formation in the Solar neighborhood.

\begin{figure}
        \centering
        \includegraphics[width=\columnwidth,angle=0]{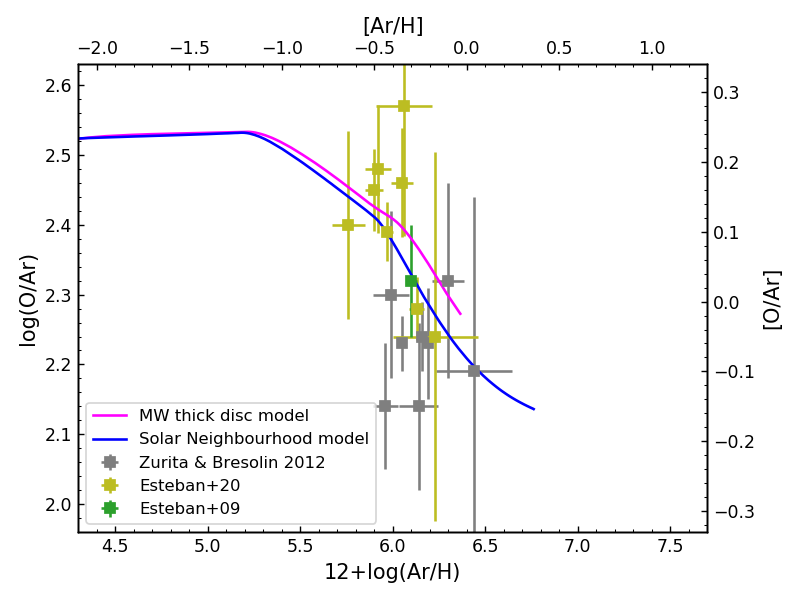}
        \caption{Same as Figure~\ref{fig:chem} but showing the M31 H{\rm II} region log(O/Ar) values as function of their 12+log(Ar/H) together with the chemical evolution tracks of \citet{Kobayashi20}. The [Ar/H] abundances and [O/Ar] ratios with respect to the solar values are marked on the top and right axis, respectively.}
        \label{fig:chem_h2}
\end{figure}

\subsection{Fresh infall of gas: oxygen and argon abundances of HII regions in the M31 disc}
\label{sect:h2}

A useful insight on the chemical evolution of the M31 discs is given by the properties of the younger generation of stars with respect to the older low-extinction PNe in the M31 thicker disc. We thus explore the log(O/Ar) vs. 12 + log(Ar/H) distribution (Figure~\ref{fig:chem_h2}) measured for the H{\rm II} regions in the M~31 thin disc (see Section~\ref{sect:oar_grad}). Similar to the M31 PNe, the H{\rm II} regions follow a decreasing trend in log(O/Ar) with 12 + log(Ar/H) with a significant scatter. When we compare the H{\rm II} region log(O/Ar) distribution with the MW thick/thin chemical enrichment models, we see that the H{\rm II} region values do not cluster near the end of the track for the MW thin disc, at high [Ar/H] values. At a given argon abundance, their large scatter supports the hypothesis that their distribution in the log(O/Ar) vs. 12 + log(Ar/H) plane is due to differences in present day abundances. The range of log(O/Ar) values, between 2.15 to 2.6, at argon abundance 12+log(Ar/H)$ \simeq 6.0$ indicates spatial variations of the abundances in the M31 gas phase of the thin disc, due to different degree of mixing of metal poor gas with the enriched ISM in the disc, and also with the starburst in the $10-17$ kpc region \citep{Kang09}. The spatial variations are plausibly linked with the recent production of oxygen in the starburst and rainfall of gas over the M31 disc, which may be related to the extra planar H{\rm I} gas \citep{Westmeier08}. 

\subsection{Constraints on the chemical enrichment of the thin disc in M31 with PN abundances}
\label{sect:Hepne}

In this section we investigate the differences in the chemical enrichment of the high-extinction PNe in M~31, which are tracing the kinematically  thin disc of M~31 \citepalias{Bh+19b}, with respect to the low-extinction older PNe.  We note that high-extinction PNe occupy regions of the log(O/Ar) vs. 12+log(Ar/H) plane that, even if with some overlaps, are somewhat different from the low-extinction PNe, see Figure~\ref{fig:arh_oh} right plot. We find that the decrease for the mean log(O/Ar)  vs. 12+log(Ar/H) is linear for the low-extinction PNe. The mean log(O/Ar) vs. 12+log(Ar/H) for the  high-extinction PNe are larger than those for the low extinction PNe in most bins, but for the 12+log(Ar/H) = 6.1 -- 6.3 range (3 bins), where they are lower.

\begin{figure}
        \centering
        \includegraphics[width=\columnwidth,angle=0]{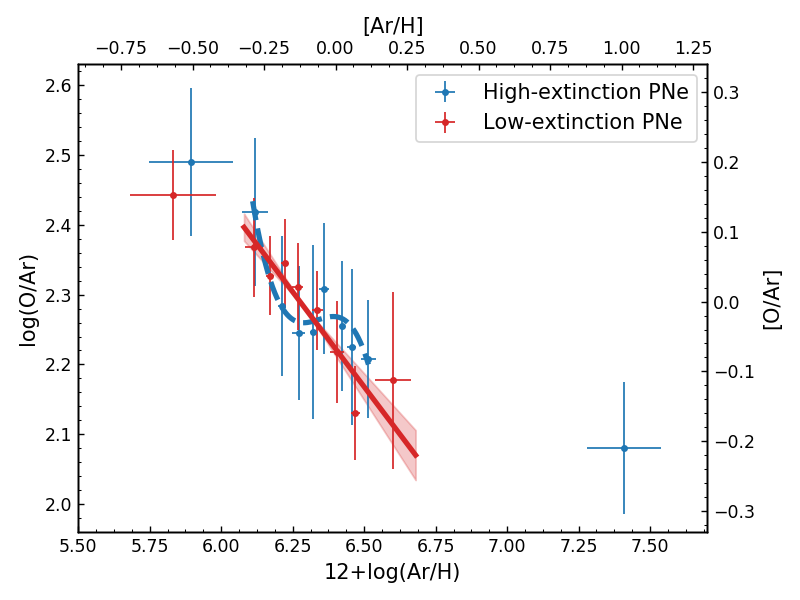}
        \caption{{Same as Figure~\ref{fig:chem} showing the mean log(O/Ar) values of the high- (blue) and low- (red) extinction PNe but now also showing the linear fit with uncertainty shaded for the low-extinction sample. For the high-extinction PNe, the dashed blue line shows a third order polynomial curve which better fits the lower mean log(O/Ar) values of the high-extinction PNe in the 12+log(Ar/H) $= 6.1 -6.3$ range}. The [Ar/H] abundances and [O/Ar] ratios with respect to the solar values are marked on the top and right axis, respectively. }
        \label{fig:loop1}
\end{figure}

 We compare the linear fit to the mean log(O/Ar) values in the 12+log(Ar/H) = 6 -- 6.7 range, for the high- and low-extinction PNe in the M~31 disc, and validate them statistically. The high- and low-extinction PNe can be fitted by linear functions with a Bayesian Information Criterion\footnote{It is a criterion for model selection among a finite set of models based on bayesian statistics. Overfitting may result from adding parameters to increase the likelihood of a function. BIC introduces a penalty term for the number of parameters in the model in order to avoid overfitting.} \citep[BIC;][]{Schwarz78} values of -50.42 and -48.99, respectively. The linear fit to the low-extinction PNe (slope~$=-0.54\pm0.09$; intercept~$=5.7\pm0.58$) is the best-fit, see Figure~\ref{fig:loop1}, with higher order functions not producing better fits (giving higher BIC values instead). A third order polynomial does however provide a better fit to the high-extinction PNe (BIC=-56.17) as shown in Figure~\ref{fig:loop1}, thus validating the deviation from simple linear decrease. 

If both the high- and low-extinction PNe were following the same enrichment history, the high-extinction PNe, which have an age of $\sim2.5$~Gyr or younger (\citetalias{Bh+19b}), should populate low log(O/Ar)- higher argon abundances region of the evolutionary track, see Section~\ref{sect:olddisc} and \ref{sect:h2}. The existence of high-extinction younger PNe with high log(O/Ar) values at relative lower argon abundances support a secondary  infall event with less chemically evolved gas. 


\section{Constraints on chemical evolution and the formation history of the M~31 disc}
\label{sect:disc2}

As illustrated in Section~\ref{sect:arg_dep}, galactic chemical evolution models of the MW thick disc and solar neighbourhood \citep{Kobayashi20} have a near linear decrease of log(Ar/O) values in the 12+log(Ar/H) = 6 -- 6.7 range (see Figure~\ref{fig:chem}). Deviations from a simple linearly decreasing chemical evolution track may be related to either quenching or a secondary infall of gas onto a galaxy which then causes a modification of the ISM chemical abundances (see Section~\ref{sect:loop} for details and \citealt{Matteucci21} for a review). 

\begin{figure*}
        \centering
        \includegraphics[width=0.65\columnwidth,angle=0]{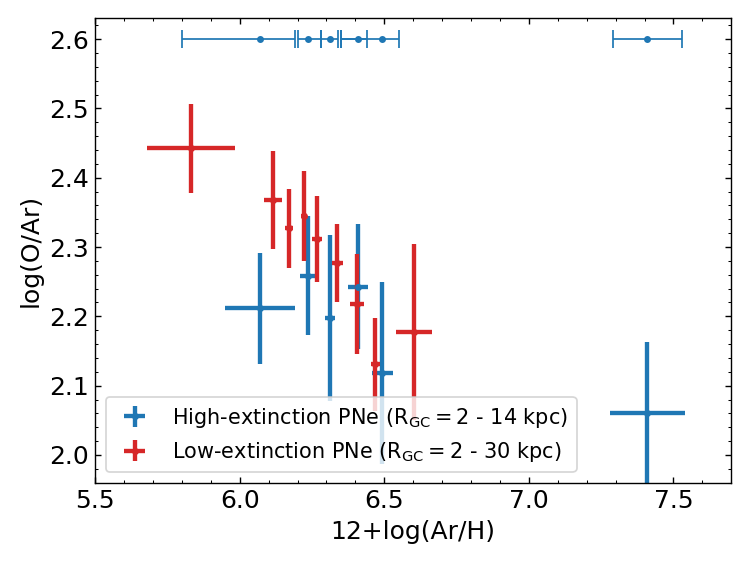}
        \includegraphics[width=0.65\columnwidth,angle=0]{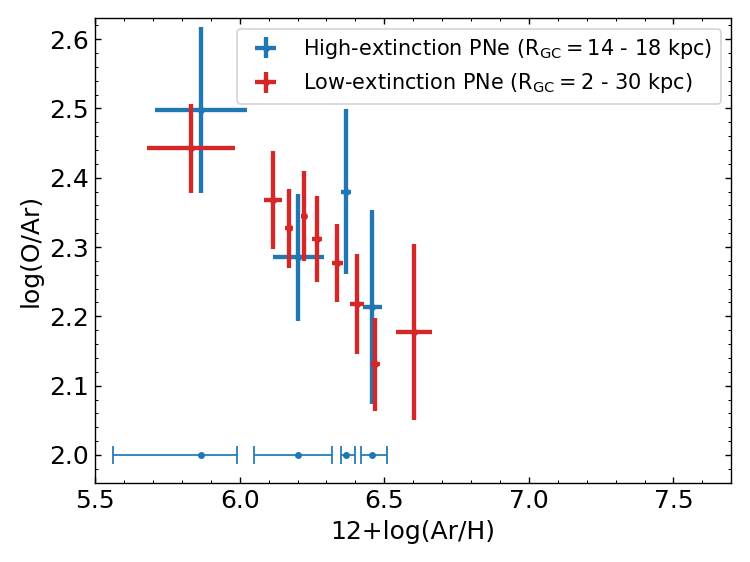}
        \includegraphics[width=0.65\columnwidth,angle=0]{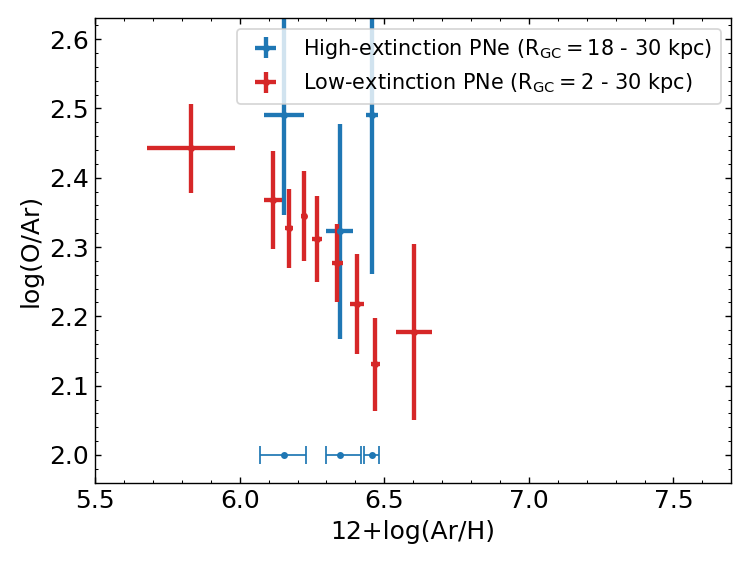}
        \caption{Same as Figure~\ref{fig:chem} showing the mean log(O/Ar) values of the high- (blue) and low- (red) extinction PNe but showing [Left] only those high-extinction PNe in the inner dsc (R$\rm_{GC}<14$~kpc), [Middle] starburst region ($\rm 14 \leq R_{GC} \leq 18$~kpc), and [Right] outer disc (R$\rm_{GC}>18$~kpc).
        The blue error-bars parallel to the x-axis indicate the width of the bins in 12+log(A/H) for the high extinction (blue) PN subsamples in the different annuli. }
        \label{fig:loop}
\end{figure*}

\subsection{Regions of homogeneous chemical evolution in M31 discs}\label{sect:annuli}
The chemical enrichment models do not include the effects of radial metallicity gradients. Therefore, as the high-extinction PNe have i) a steeper argon abundance gradient (\citetalias{Bhattacharya22}) and ii) have distinct behaviour of the log(O/Ar) values in three radial ranges, as described in Section~\ref{sect:oar_grad}, the steps prior to the chemical modeling include the identification of the disc regions with nearly flat log(O/Ar) radial gradients.

In Figure~\ref{fig:loop} we show the distribution of the high-extinction PNe in the three radial ranges, inner disc (R$\rm_{GC}<14$~kpc) [Left],  starburst region ($\rm 14 \leq R_{GC} \leq 18$~kpc) [Middle], and outer disc (R$\rm_{GC}>18$~kpc)  [Right], with respect to the low extinction PN values for the entire M31 disc. While in Section~\ref{sect:Hepne} the high extinction PNe have larger log(O/Ar) values than the low extinction PNe, we find that the high extinction, younger, PNe in the inner disc cluster towards the end, and below, the low extinction PNe distribution at relatively high [Ar/H], see left panel
of Fig.~\ref{fig:loop}. Considering their younger age, they are consistent with the chemical evolution only if also there is a dilution of the ISM by infall of metal poorer gas. In the middle panel of Figure~\ref{fig:loop}, the high extinction PNe show a large range in argon abundances for a 2.5 Gyr and younger evolution. Those points on the top of the low-extinction PNe distribution may be consistent with the strong starburst located in this disc region \citep{Kang09} and a range of initial gas mixing. In the right panel of Figure~\ref{fig:loop}, the high log(O/Ar) values at relatively low argon abundances for the younger high extinction PNe, with respect to the low extinction PNe distribution, indicate a higher star formation efficiency in those outer regions, i.e. shorter timescales in the thin disc with respect to the thicker disc, with the progenitor of these younger PNe being formed mostly out of the less enriched gas. 

\subsection{Imprint of a secondary gas infall on the oxygen and argon abundances in the inner (R$\rm_{GC}<14$~kpc) M31 thin disc from a chemical evolution model}
\label{sect:loop}

To understand the chemical evolution of the younger disc in M~31 within 14 kpc,  we first review the arguments available in the literature used to describe the chemical evolution of the MW discs. Two infall-model scenarios were proposed to explain the distribution of the MW stars in the [$\alpha$/Fe] (or [Mg/Fe]) vs. [Fe/H] plane, in different radial ranges, see models by \citet{Grisoni17,Spitoni19,Spitoni21}. According to these models, the MW thick disc stars form rapidly with high star-formation efficiency which leads to high [$\alpha$/Fe] values for its stars. Eventually star-formation is suppressed as gas is depleted in the MW thick disc. After some time  (t$\rm_{max}$), further infall of  gas, either pristine or pre-enriched, occurs which dilutes the available ISM and then triggers star-formation, at a lower efficiency. This may lead to the formation of loops in the galactic chemical evolution tracks in the [$\alpha$/Fe] (or [Mg/Fe]) vs. [Fe/H] plane \citep{Matteucci21}. Thin disc stars in the MW are formed initially at a lower metallicity which then increases with time. In the the MW thin disc, stars have lower [$\alpha$/Fe] values than thick disc stars because of the diminished star formation efficiency. The time interval t$\rm_{max}$, the amount of accreted gas, and whether the latter is pristine or pre-enriched, drives the extent of these loops in the galactic chemical evolution tracks. Larger loops can be present for larger t$\rm_{max}$ values \citep[see Figure 14 in][]{Spitoni19} and/or larger amount of pristine gas. The loop also becomes narrower if the infalling gas is pre-enriched \citep[][]{Spitoni21}.

As illustrated in Section~\ref{sect:arg_dep} and \ref{sect:annuli}, the oxygen and argon abundance distribution and ratios for young high extinction PNe  support a secondary event of gas accretion.
Hence we construct new, galactic chemical evolution models for the inner disc regions (R$\rm_{GC}<14$~kpc) with two gas inflow phases for the M31 disc, using the same chemical evolution code by \citet{Kobayashi20}. The models are constrained using the star-formation history and metallicity distribution function measured for the M~31 disc within the PHAT footprint (R$\rm_{GC}\simeq 18$~kpc) using isochrone fitting to the observed RGB stars \citep{wil17}. See Appendix~\ref{sect:chem-evol-W17} for a detailed illustration of these models.

\begin{figure*}
        \centering
        \includegraphics[width=\columnwidth,angle=0]{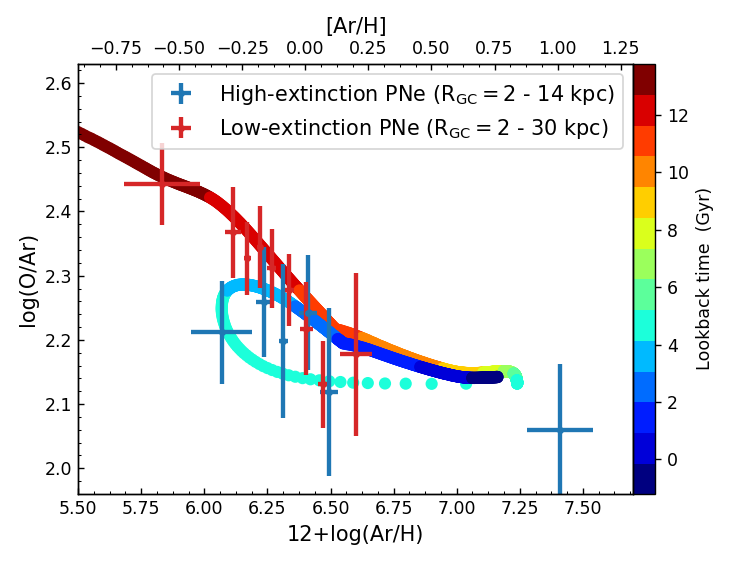}
        \includegraphics[width=\columnwidth,angle=0]{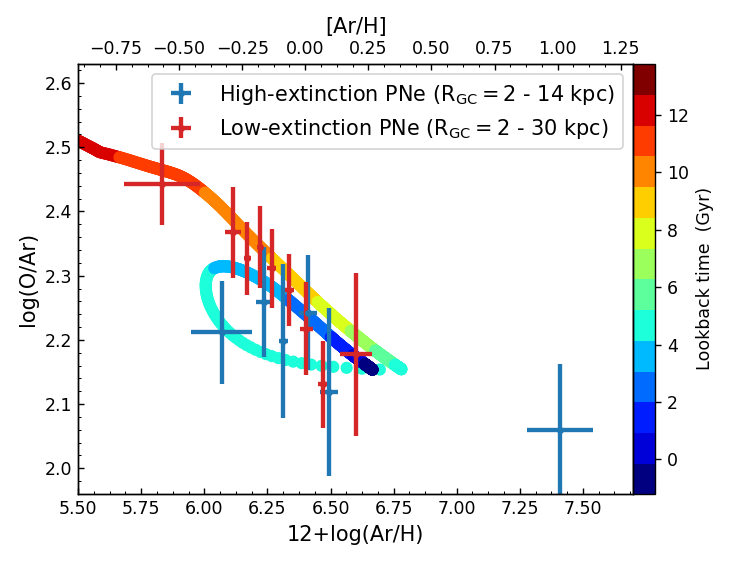}
        \caption{The Figure shows the mean log(O/Ar) values of the low- (red) extinction PNe over the 2-30 kpc radial range,  the high- (blue) extinction PNe within  R$\rm_{GC}<14$~kpc and the new two-infall chemical evolution model for the M~31 disc (see Section~\ref{sect:loop} for details), coloured by lookback time: the fiducial model on the left and the alternative model with  a weaker first star formation on the right. The [Ar/H] abundances with respect to the solar values are marked on the top and right axis, respectively.  See Appendix~\ref{sect:chem-evol-W17} for more details on the chemical evolution models.}
        \label{fig:loop_model}
\end{figure*}

In these chemical evolution models, the thicker disc older stars are formed by the first gas inflow, with the infall timescale of 1 Gyr. The SFR of the first burst differs in the models though, with a weaker first burst by a factor $\approx 2.5$ for the second model, which is still within the uncertainties of the total star formation rate vs. age for the entire PHAT footprint, see \citet{wil17} and Appendix~\ref{sect:chem-evol-W17}.  The star formation rate ceases after 2 Gyr and 3 Gyr, in the fiducial and weaker first burst model, respectively.  The thin disc stars instead are formed by the second inflow, started at 9 Gyr with the infall timescale of 2 Gyr. In the two models, the SFR for the second burst differs at the 20\% level, with the fiducial model having a relatively weaker second burst. The chemical composition of both gas inflows is set to be primordial, which results in the loop in the log(O/Ar) vs. 12+log(Ar/H) plane. The star formation timescales are 10 Gyr and 2 Gyr for the thicker and thin disc stars respectively, which means that the star formation efficiency is five times higher during the thin disc formation than for the thicker disc. A small amount of gas outflow (with the timescale of 5~Gyr) is also included. These new models are shown in Figure~\ref{fig:loop_model} together with the high extinction PN oxygen and argon measurements for the M31 ISM within 14 kpc, in the log(O/Ar) vs. 12+log(Ar/H) plane, and the low extinction PNe log(O/Ar) vs. 12+log(Ar/H) measurements over the entire disc. The list of the relevant parameters for the chemical evolution models is given in Table~\ref{tab:chem_mod}.

We discuss the two chemical enrichment tracks in turn. 
In Figure~\ref{fig:loop_model}~[Left], the fiducial model described in Appendix~\ref{sect:chem-evol-W17} reaches 12+log(Ar/H) $\sim$7.25 at lookback time $\sim5$~Gyr ago, following which the star formation is suppressed in the thick disc. Infall of primordial gas occurs $\sim 5$ Gyr ago; it rapidly reduces the mean 12+log(Ar/H) value of the ISM and leads to a second burst of star-formation. At the start of the second burst of star-formation, prior to supernovae Type-Ia eruption, the log(O/Ar) value at corresponding lower metallicity increases, thereby producing the rising part of the loop (Figure~\ref{fig:loop_model}-[Left]). The loop turns over once the supernovae Type-Ia eruption kicks in, leading to a decrease of the log(O/Ar) value with increasing 12+log(Ar/H), over the past $\sim$4 Gyr. The track reaches 12+log(Ar/H) values $\simeq$ 7.2 at present times. The high argon abundance high extinction PNe are better reproduced by this fiducial model .

In Figure~\ref{fig:loop_model}~[Right], because of the weaker first star formation, the second model reaches 12+log(Ar/H) $\sim$6.7 at lookback time $\sim5$~Gyr ago. With the dilution and loop following a similar pattern as for the model in the left panel, the track reaches 12+log(Ar/H) values $\simeq$ 6.7 at present times. This chemical evolution model of the ISM in the M~31 thin/thicker disc is in good agreement with the values traced by the low-extinction PNe and the loop covers the distribution of the high extinction PNe log(O/Ar) and 12+log(Ar/H) values ar $R_{GC} < 14$ kpc. However, super-solar metallicity values traced by the high-extinction PNe are not reached in this model, and the model seems to underestimate the amount of low-metallicity gas added. Such remaining discrepancies may be addressed in a future investigation with more extended chemo-dynamical modelling. 

\begin{figure}
        \centering
        \includegraphics[width=\columnwidth,angle=0]{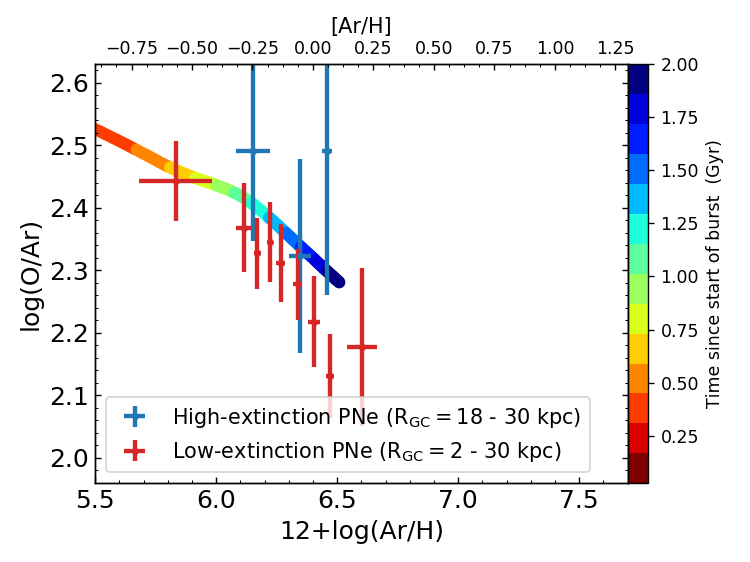}
        \caption{The plot shows the mean log(O/Ar) values of the low- (red) extinction PNe over the 2-30 kpc radial range,  the high- (blue) extinction PNe in the outer disc R$\rm_{GC}>18$~kpc. The chemical evolution track shows the chemical enrichment within approx 2 Gyr following a burst of star formation. See Appendix~\ref{sect:chem-evol-W17} for more details. }
        \label{fig:outerdisc}
\end{figure}
In Figure~\ref{fig:outerdisc} we show the chemical evolution enrichment track computed with reference to a burst of star formation from a gas infall with primordial composition and a timescale of 2 Gyr to reproduce the chemical enrichment of the thin outer disc, at $R\rm_{GC}>18$ kpc. The lookback time shows the time from the start of the gas infall. The adopted timescale is even shorter than that of the MW thick disc, and it is as short as for nearby early-type galaxies It clearly illustrates the shorter star formation timescale and the lower argon abundance of the outer thin disc in M31. 

\section{Possible gas content of the merging satellite in the M~31 disc}
\label{sect:satgas}

From the age-velocity dispersion relation in the 14-17 kpc and 17-20 kpc, in \citetalias{Bh+19b} we constrained the baryonic mass of the satellite that merged with the M31 disc. The estimated merger mass ratio from  
the age-velocity dispersion relation is $1:5$ \citep{Hopkins08, Hopkins09}, hence the baryonic mass of the satellite M$\rm_{sat} = 1.4 \times 10^{10}$~M$\rm_{\odot}$. Adopting the best-fit galaxy gas-fraction to galaxy stellar mass relation \citep[][]{Diaz20}, the estimated gas fraction for the satellite would then be $\sim$19\% , with a total mass in gas of M$\rm_{HI,sat} = 2.8 \times 10^{9}$~M$\rm_{\odot}$. We can compare the estimated gas mass brought in from the satellite with the i) mass in stars formed during the last 4.5 Gyr and ii) the amount of neutral H{\rm I} still detected in the M31 disc. 

\citet{wil17} found that the total stellar mass produced over the last $\sim2$~Gyr in the M~31 disc is M$\rm_{stellar} \sim 4.5 \times 10^{9}$~M$\rm_{\odot}$, and the present day total mass of H{\rm I} gas in the M~31 disc is M$\rm_{HI} = 4.23 \times 10^{9}$~M$\rm_{\odot}$ \citep{Chemin09} . Thus, prior to the burst of star-formation $\sim$2.5 Gyr ago \citep{bernard15,wil17}, a gas mass of M$\rm_{gas,2~Gyr} \simeq 9 \times 10^{9}$~M$\rm_{\odot}$ must have been available in the M~31 disc. This amount is much larger than the estimated amount brought in by the merging satellite. Thus residual gas had to be  still present in the pre-merger M~31 disc, which was then diluted by $\Delta [Ar/H] \simeq -0.5$ dex as observed in the argon abundance distribution. The high extinction PNe were formed out of this diluted ISM within 14 kpc, as illustrated in Figure~\ref{fig:loop_model}, and mostly from satellite gas at $R_{GC} > 18$ kpc, see Figure~\ref{fig:outerdisc} . 

From the mass-metallicity relation \citep[e.g.][]{Zahid17}, the gas from the merging satellite galaxy would be pre-enriched, but at lower metallicity values than the ISM of the M31 pre-merger disc. A galaxy of mass equal to M$\rm_{sat}$ is expected to have a metallicity $\sim-0.5$ dex. The metallicity distribution of the RGB stars in the GSS from \citet{Conn16}, and \citet{Cohen18} provide evidence for a metal poor population of stars with [Fe/H] in the range $[-1.0:-0.2]$ dex. Then a second infall of gas related to the wet merger of the satellite onto the M31 disc would dilute the previously enriched ISM in the M31 pre-merger disc. The star burst triggered by the newly acquired gas would reproduce the lower [Ar/H] values and the log(O/Ar) distribution of the high-extinction PNe, as discussed in Section~\ref{sect:loop}. The lower metallicity of the gas in the satellite, estimated either from the mass-metallicity relation or the observation of the GSS stars, is consistent with extent of the ``loop'' towards lower [Ar/H] abundances in the chemical evolution models shown in Figure~\ref{fig:loop_model}.


\section{Conclusions}
\label{sect:future}
We use the largest sample of PNe in the M~31 disc with oxygen and argon measurements from \citetalias{Bhattacharya22}, as well as archival measurements of these elements for HII regions.  We compare their distributions with chemical evolution models for the MW thick disc, the Solar neighborhood and  chemical enrichment models for the M31 discs with secondary infall phases. Our main conclusions are that the log(O/Ar) vs. argon abundance distributions for the nebular emissions of the stellar evolution phases (e.g. HII regions and PNe) are valid alternatives to the [$\alpha$/Fe] vs. [Fe/H] diagrams that have been widely used to constrain the star formation efficiency in the MW. By using this new approach, we determined a different chemical evolution history for the thin and thicker discs in the Andromeda galaxy, with the thin disc in M31 having been affected by a wet merger. In detail, our conclusions can be further illustrated as follows:
\begin{itemize}
    \item The oxygen-to-argon ratio, log(O/Ar), values of PNe are high at low argon abundances and monotonically decrease to lower values at higher argon abundances, consistent with predictions of the ISM properties from galactic chemical evolution models. This is the first vivid example that galactic chemical enrichment is imprinted in the log(O/Ar) vs. Ar plane.
    \item The log(O/Ar) vs. argon abundance distribution for the low-extinction, 4.5 Gyr and older, PNe associated with the M~31 thicker disc supports a chemical evolution where the thicker disc forms from infalling gas with primordial composition at early times and reaches a high argon abundance  similar to that of the Solar neighborhood in the MW (12+log(Ar/H)$\simeq 6.7$).
    \item  The log(O/Ar) vs. 12+log(Ar/H) distribution for the high-extinction, 2.5 Gyr and younger, PNe associated with the thin disc of M31 overlaps with that of the low extinction PNe, with quantifiable differences. Three regions are identified in the M31 disc: inner disc, starburst region and outer disc. 
    \item In the inner disc ($R_{GC} < 14$ kpc),  the high-extinction, 2.5 Gyr and younger, PNe form after a secondary event with infall of metal poor  gas which mixed with the pre-enriched ISM in the pre-merger M31 disc. This is in agreement with the two-infall galactic chemical evolution model described in Section~\ref{sect:loop} and Appendix~\ref{sect:chem-evol-W17}. In the starburst region ($14 \leq R_{GC} \leq 18$ kpc), the high-extinction, 2.5 Gyr and younger, PNe display a range of initial gas mixing and high log(O/Ar) values, consistent with the strong starburst located in this region of the M31 disc (see \citealt{Kang09}). In the $R_{GC} > 18$ outer disc, the high extinction PNe progeny are formed mostly in a burst of star formation $\simeq$ 2 Gyr ago from satellite less chemically enriched gas. 
    \item The log(O/Ar) vs. 12+log(Ar/H) distribution for the H{\rm II} regions in M31 has a large scatter at fixed argon abundance 12+log(Ar/H)$\simeq 6.2$. This spread is consistent with spatial variations and different degree of mixing, which is reasonably linked to current rainfall of metal poor gas from extra planar HI, part of which may have also originated from the satellite. 
    \item {In contrast to the chemical enrichment history and structure of the thin disc in the MW, the thin disc in M~31 is less radially extended, formed stars more recently and at a higher star-formation efficiency, and had a faster chemical enrichment timescale than the thicker disc in M31. }

\end{itemize}
The next steps of this investigation will explore the kinematical and chemical abundance properties of the PNe associated with the substructures in the outer disc and inner halo of M31. 

\begin{acknowledgements}
     MAR and SB thank the European Southern Observatory (ESO), Garching, Germany for supporting SB's visit through the 2021 ESO SSDF. MAR, SB and OG are grateful for the hospitality of the Mount Stromlo
Observatory and the Australian National University (ANU). MAR and OG
thank the Research School of Astronomy and Astrophysics at ANU for support
through their Distinguished Visitor Program. This work was supported by the
DAAD under the Australia-Germany joint research program with funds from
the German Federal Ministry for Education and Research. SB is funded by the INSPIRE Faculty award (DST/INSPIRE/04/2020/002224), Department of Science and Technology (DST), Government of India. CK acknowledge funding 
    from the UK Science and Technology Facility Council through grant ST/R000905/1 and ST/V000632/1. Based on observations obtained at the MMT Observatory, a joint facility of the Smithsonian Institution and the University of Arizona. Based on observations obtained with MegaPrime/MegaCam, a joint project of CFHT and CEA/DAPNIA, at the Canada-France-Hawaii Telescope (CFHT). This research made use of Astropy-- a community-developed core Python package for Astronomy \citep{Rob13}, SciPy \citep{scipy}, NumPy \citep{numpy} and Matplotlib \citep{matplotlib}. This research also made use of NASA’s Astrophysics Data System (ADS\footnote{\url{https://ui.adsabs.harvard.edu}}).
\end{acknowledgements}


\bibliographystyle{aa} 
\bibliography{ref_pne.bib}

\begin{thebibliography}{72}
\expandafter\ifx\csname natexlab\endcsname\relax\def\natexlab#1{#1}\fi

\bibitem[{{Aniyan} {et~al.}(2018){Aniyan}, {Freeman}, {Arnaboldi}, {Gerhard},
  {Coccato}, {Fabricius}, {Kuijken}, {Merrifield}, \& {Ponomareva}}]{Aniyan18}
{Aniyan}, S., {Freeman}, K.~C., {Arnaboldi}, M., {et~al.} 2018, \mnras, 476,
  1909

\bibitem[{{Aniyan} {et~al.}(2021){Aniyan}, {Ponomareva}, {Freeman},
  {Arnaboldi}, {Gerhard}, {Coccato}, {Kuijken}, \& {Merrifield}}]{Aniyan21}
{Aniyan}, S., {Ponomareva}, A.~A., {Freeman}, K.~C., {et~al.} 2021, \mnras,
  500, 3579

\bibitem[{{Astropy Collaboration} {et~al.}(2013){Astropy Collaboration},
  {Robitaille}, {Tollerud}, {Greenfield}, {Droettboom}, {Bray}, {Aldcroft},
  {Davis}, {Ginsburg}, {Price-Whelan}, {Kerzendorf}, {Conley}, {Crighton},
  {Barbary}, {Muna}, {Ferguson}, {Grollier}, {Parikh}, {Nair}, {Unther},
  {Deil}, {Woillez}, {Conseil}, {Kramer}, {Turner}, {Singer}, {Fox}, {Weaver},
  {Zabalza}, {Edwards}, {Azalee Bostroem}, {Burke}, {Casey}, {Crawford},
  {Dencheva}, {Ely}, {Jenness}, {Labrie}, {Lim}, {Pierfederici}, {Pontzen},
  {Ptak}, {Refsdal}, {Servillat}, \& {Streicher}}]{Rob13}
{Astropy Collaboration}, {Robitaille}, T.~P., {Tollerud}, E.~J., {et~al.} 2013,
  \aap, 558, A33

\bibitem[{{Belokurov} {et~al.}(2018){Belokurov}, {Erkal}, {Evans}, {Koposov},
  \& {Deason}}]{Belokurov18}
{Belokurov}, V., {Erkal}, D., {Evans}, N.~W., {Koposov}, S.~E., \& {Deason},
  A.~J. 2018, \mnras, 478, 611

\bibitem[{{Belokurov} {et~al.}(2020){Belokurov}, {Sanders}, {Fattahi}, {Smith},
  {Deason}, {Evans}, \& {Grand}}]{Belokurov20}
{Belokurov}, V., {Sanders}, J.~L., {Fattahi}, A., {et~al.} 2020, \mnras, 494,
  3880

\bibitem[{{Bernard} {et~al.}(2015){Bernard}, {Ferguson}, {Richardson}, {Irwin},
  {Barker}, {Hidalgo}, {Aparicio}, {Chapman}, {Ibata}, {Lewis}, {McConnachie},
  \& {Tanvir}}]{bernard15}
{Bernard}, E.~J., {Ferguson}, A. M.~N., {Richardson}, J.~C., {et~al.} 2015,
  \mnras, 446, 2789

\bibitem[{{Bhattacharya} {et~al.}(2022){Bhattacharya}, {Arnaboldi}, {Caldwell},
  {Gerhard}, {Kobayashi}, {Hartke}, {Freeman}, {McConnachie}, \&
  {Guhathakurta}}]{Bhattacharya22}
{Bhattacharya}, S., {Arnaboldi}, M., {Caldwell}, N., {et~al.} 2022, arXiv
  e-prints, arXiv:2203.06428

\bibitem[{{Bhattacharya} {et~al.}(2021){Bhattacharya}, {Arnaboldi}, {Gerhard},
  {McConnachie}, {Caldwell}, {Hartke}, \& {Freeman}}]{Bh21}
{Bhattacharya}, S., {Arnaboldi}, M., {Gerhard}, O., {et~al.} 2021, \aap, 647,
  A130

\bibitem[{{Bhattacharya} {et~al.}(2019{\natexlab{a}}){Bhattacharya},
  {Arnaboldi}, {{\GG{20020530}}Hartke}, {Gerhard}, {Comte}, {McConnachie}, \&
  {Caldwell}}]{Bh+19}
{Bhattacharya}, S., {Arnaboldi}, M., {{\GG{20020530}}Hartke}, J., {et~al.}
  2019{\natexlab{a}}, A\&A, 624, A132

\bibitem[{{Bhattacharya} {et~al.}(2019{\natexlab{b}}){Bhattacharya},
  {Arnaboldi}, {{\GG{20020630}}Caldwell}, {Gerhard}, {Bla{\~n}a},
  {McConnachie}, {Hartke}, {Guhathakurta}, {Pulsoni}, \& {Freeman}}]{Bh+19b}
{Bhattacharya}, S., {Arnaboldi}, M., {{\GG{20020630}}Caldwell}, N., {et~al.}
  2019{\natexlab{b}}, \aap, 631, A56

\bibitem[{{Bland-Hawthorn} \& {Gerhard}(2016)}]{bhg16}
{Bland-Hawthorn}, J. \& {Gerhard}, O. 2016, Annual Review of Astronomy and
  Astrophysics, 54, 529

\bibitem[{{Chemin} {et~al.}(2009){Chemin}, {Carignan}, \& {Foster}}]{Chemin09}
{Chemin}, L., {Carignan}, C., \& {Foster}, T. 2009, \apj, 705, 1395

\bibitem[{{Chiappini} {et~al.}(1997){Chiappini}, {Matteucci}, \&
  {Gratton}}]{Chiappini97}
{Chiappini}, C., {Matteucci}, F., \& {Gratton}, R. 1997, \apj, 477, 765

\bibitem[{{Cohen} {et~al.}(2018){Cohen}, {Kalirai}, {Gilbert}, {Guhathakurta},
  {Peeples}, {Lehner}, {Brown}, {Bianchi}, {Barger}, \& {O'Meara}}]{Cohen18}
{Cohen}, R.~E., {Kalirai}, J.~S., {Gilbert}, K.~M., {et~al.} 2018, \aj, 156,
  230

\bibitem[{{Comer{\'o}n} {et~al.}(2019){Comer{\'o}n}, {Salo}, {Knapen}, \&
  {Peletier}}]{Comeron19}
{Comer{\'o}n}, S., {Salo}, H., {Knapen}, J.~H., \& {Peletier}, R.~F. 2019,
  \aap, 623, A89

\bibitem[{{Conn} {et~al.}(2016){Conn}, {McMonigal}, {Bate}, {Lewis}, {Ibata},
  {Martin}, {McConnachie}, {Ferguson}, {Irwin}, {Elahi}, {Venn}, \&
  {Mackey}}]{Conn16}
{Conn}, A.~R., {McMonigal}, B., {Bate}, N.~F., {et~al.} 2016, \mnras, 458, 3282

\bibitem[{{Cortesi} {et~al.}(2013){Cortesi}, {Arnaboldi}, {Coccato},
  {Merrifield}, {Gerhard}, {Bamford}, {Romanowsky}, {Napolitano}, {Douglas},
  {Kuijken}, {Capaccioli}, {Freeman}, {Chies-Santos}, \& {Pota}}]{cor13}
{Cortesi}, A., {Arnaboldi}, M., {Coccato}, L., {et~al.} 2013, \aap, 549, A115

\bibitem[{{Delgado-Inglada} {et~al.}(2014){Delgado-Inglada}, {Morisset}, \&
  {Stasi{\'n}ska}}]{Delgado14}
{Delgado-Inglada}, G., {Morisset}, C., \& {Stasi{\'n}ska}, G. 2014, \mnras,
  440, 536

\bibitem[{{Delgado-Inglada} {et~al.}(2015){Delgado-Inglada}, {Rodr{\'\i}guez},
  {Peimbert}, {Stasi{\'n}ska}, \& {Morisset}}]{Delgado-Inglada15}
{Delgado-Inglada}, G., {Rodr{\'\i}guez}, M., {Peimbert}, M., {Stasi{\'n}ska},
  G., \& {Morisset}, C. 2015, \mnras, 449, 1797

\bibitem[{{D{\'\i}az-Garc{\'\i}a} \& {Knapen}(2020)}]{Diaz20}
{D{\'\i}az-Garc{\'\i}a}, S. \& {Knapen}, J.~H. 2020, \aap, 635, A197

\bibitem[{{Dorman} {et~al.}(2015){Dorman}, {Guhathakurta}, {Seth}, {Weisz},
  {Bell}, {Dalcanton}, {Gilbert}, {Hamren}, {Lewis}, {Skillman}, {Toloba}, \&
  {Williams}}]{dorman15}
{Dorman}, C.~E., {Guhathakurta}, P., {Seth}, A.~C., {et~al.} 2015, \apj, 803,
  24

\bibitem[{{D'Souza} \& {Bell}(2018)}]{DSouza18b}
{D'Souza}, R. \& {Bell}, E.~F. 2018, \mnras, 474, 5300

\bibitem[{{Esteban} {et~al.}(2020){Esteban}, {Bresolin}, {Garc{\'\i}a-Rojas},
  \& {Toribio San Cipriano}}]{Esteban20}
{Esteban}, C., {Bresolin}, F., {Garc{\'\i}a-Rojas}, J., \& {Toribio San
  Cipriano}, L. 2020, \mnras, 491, 2137

\bibitem[{{Esteban} {et~al.}(2009){Esteban}, {Bresolin}, {Peimbert},
  {Garc{\'\i}a-Rojas}, {Peimbert}, \& {Mesa-Delgado}}]{Esteban09}
{Esteban}, C., {Bresolin}, F., {Peimbert}, M., {et~al.} 2009, \apj, 700, 654

\bibitem[{{Fabricant} {et~al.}(2005){Fabricant}, {Fata}, {Roll}, {Hertz},
  {Caldwell}, {Gauron}, {Geary}, {McLeod}, {Szentgyorgyi}, {Zajac}, {Kurtz},
  {Barberis}, {Bergner}, {Brown}, {Conroy}, {Eng}, {Geller}, {Goddard},
  {Honsa}, {Mueller}, {Mink}, {Ordway}, {Tokarz}, {Woods}, {Wyatt}, {Epps}, \&
  {Dell'Antonio}}]{fab05}
{Fabricant}, D., {Fata}, R., {Roll}, J., {et~al.} 2005, Publications of the
  Astronomical Society of the Pacific, 117, 1411

\bibitem[{{Garc{\'\i}a-Hern{\'a}ndez} \& {G{\'o}rny}(2014)}]{GarciaHernandez14}
{Garc{\'\i}a-Hern{\'a}ndez}, D.~A. \& {G{\'o}rny}, S.~K. 2014, \aap, 567, A12

\bibitem[{{Garc{\'\i}a-Hern{\'a}ndez}
  {et~al.}(2016){Garc{\'\i}a-Hern{\'a}ndez}, {Ventura}, {Delgado-Inglada},
  {Dell'Agli}, {Di Criscienzo}, \& {Yag{\"u}e}}]{Garcia-Hernandez16}
{Garc{\'\i}a-Hern{\'a}ndez}, D.~A., {Ventura}, P., {Delgado-Inglada}, G.,
  {et~al.} 2016, \mnras, 458, L118

\bibitem[{{Gilmore} \& {Reid}(1983)}]{Gilmore83}
{Gilmore}, G. \& {Reid}, N. 1983, \mnras, 202, 1025

\bibitem[{{Grisoni} {et~al.}(2017){Grisoni}, {Spitoni}, {Matteucci},
  {Recio-Blanco}, {de Laverny}, {Hayden}, {Mikolaitis}, \&
  {Worley}}]{Grisoni17}
{Grisoni}, V., {Spitoni}, E., {Matteucci}, F., {et~al.} 2017, \mnras, 472, 3637

\bibitem[{{Hammer} {et~al.}(2018){Hammer}, {Yang}, {Wang}, {Ibata}, {Flores},
  \& {Puech}}]{ham18}
{Hammer}, F., {Yang}, Y.~B., {Wang}, J.~L., {et~al.} 2018, \mnras, 475, 2754

\bibitem[{{Hartke} {et~al.}(2022){Hartke}, {Arnaboldi}, {Gerhard}, {Coccato},
  {Merrifield}, {Kuijken}, {Pulsoni}, {Agnello}, {Bhattacharya}, {Spiniello},
  {Cortesi}, {Freeman}, {Napolitano}, \& {Romanowsky}}]{Hartke22}
{Hartke}, J., {Arnaboldi}, M., {Gerhard}, O., {et~al.} 2022, \aap, 663, A12

\bibitem[{{Hayden} {et~al.}(2015){Hayden}, {Bovy}, {Holtzman}, {Nidever},
  {Bird}, {Weinberg}, {Andrews}, {Majewski}, {Allende Prieto}, {Anders},
  {Beers}, {Bizyaev}, {Chiappini}, {Cunha}, {Frinchaboy},
  {Garc{\'\i}a-Her{\'n}andez}, {Garc{\'\i}a P{\'e}rez}, {Girardi}, {Harding},
  {Hearty}, {Johnson}, {M{\'e}sz{\'a}ros}, {Minchev}, {O'Connell}, {Pan},
  {Robin}, {Schiavon}, {Schneider}, {Schultheis}, {Shetrone}, {Skrutskie},
  {Steinmetz}, {Smith}, {Wilson}, {Zamora}, \& {Zasowski}}]{Hayden15}
{Hayden}, M.~R., {Bovy}, J., {Holtzman}, J.~A., {et~al.} 2015, \apj, 808, 132

\bibitem[{{Haywood} {et~al.}(2013){Haywood}, {Di Matteo}, {Lehnert}, {Katz}, \&
  {G{\'o}mez}}]{Haywood13}
{Haywood}, M., {Di Matteo}, P., {Lehnert}, M.~D., {Katz}, D., \& {G{\'o}mez},
  A. 2013, \aap, 560, A109

\bibitem[{{Helmi} {et~al.}(2018){Helmi}, {Babusiaux}, {Koppelman}, {Massari},
  {Veljanoski}, \& {Brown}}]{Helmi18}
{Helmi}, A., {Babusiaux}, C., {Koppelman}, H.~H., {et~al.} 2018, \nat, 563, 85

\bibitem[{{Hopkins} {et~al.}(2009){Hopkins}, {Cox}, {Younger}, \&
  {Hernquist}}]{Hopkins09}
{Hopkins}, P.~F., {Cox}, T.~J., {Younger}, J.~D., \& {Hernquist}, L. 2009,
  \apj, 691, 1168

\bibitem[{{Hopkins} {et~al.}(2008){Hopkins}, {Hernquist}, {Cox}, {Younger}, \&
  {Besla}}]{Hopkins08}
{Hopkins}, P.~F., {Hernquist}, L., {Cox}, T.~J., {Younger}, J.~D., \& {Besla},
  G. 2008, \apj, 688, 757

\bibitem[{Hunter(2007)}]{matplotlib}
Hunter, J.~D. 2007, Computing In Science \& Engineering, 9, 90

\bibitem[{{Ibata} {et~al.}(2001){Ibata}, {Irwin}, {Lewis}, {Ferguson}, \&
  {Tanvir}}]{Ibata01}
{Ibata}, R., {Irwin}, M., {Lewis}, G., {Ferguson}, A. M.~N., \& {Tanvir}, N.
  2001, \nat, 412, 49

\bibitem[{{Kang} {et~al.}(2009){Kang}, {Bianchi}, \& {Rey}}]{Kang09}
{Kang}, Y., {Bianchi}, L., \& {Rey}, S.-C. 2009, \apj, 703, 614

\bibitem[{{Kobayashi} {et~al.}(2020{\natexlab{a}}){Kobayashi}, {Karakas}, \&
  {Lugaro}}]{Kobayashi20}
{Kobayashi}, C., {Karakas}, A.~I., \& {Lugaro}, M. 2020{\natexlab{a}}, \apj,
  900, 179

\bibitem[{{Kobayashi} {et~al.}(2020{\natexlab{b}}){Kobayashi}, {Leung}, \&
  {Nomoto}}]{Kobayashi20b}
{Kobayashi}, C., {Leung}, S.-C., \& {Nomoto}, K. 2020{\natexlab{b}}, \apj, 895,
  138

\bibitem[{{Kobayashi} \& {Nakasato}(2011)}]{Kobayashi11}
{Kobayashi}, C. \& {Nakasato}, N. 2011, \apj, 729, 16

\bibitem[{{Kobayashi} \& {Nomoto}(2009)}]{Kobayashi09}
{Kobayashi}, C. \& {Nomoto}, K. 2009, \apj, 707, 1466

\bibitem[{{Kwitter} \& {Henry}(2021)}]{Kwitter21}
{Kwitter}, K.~B. \& {Henry}, R.~B.~C. 2021, arXiv e-prints, arXiv:2110.13993

\bibitem[{{Kwitter} {et~al.}(2012){Kwitter}, {Lehman}, {Balick}, \&
  {Henry}}]{Kw12}
{Kwitter}, K.~B., {Lehman}, E. M.~M., {Balick}, B., \& {Henry}, R.~B.~C. 2012,
  \apj, 753, 12

\bibitem[{{Maciel} \& {Koppen}(1994)}]{Maciel94}
{Maciel}, W.~J. \& {Koppen}, J. 1994, \aap, 282, 436

\bibitem[{{Magrini} {et~al.}(2016){Magrini}, {Coccato}, {Stanghellini},
  {Casasola}, \& {Galli}}]{Magrini16}
{Magrini}, L., {Coccato}, L., {Stanghellini}, L., {Casasola}, V., \& {Galli},
  D. 2016, \aap, 588, A91

\bibitem[{{Matteucci}(2021)}]{Matteucci21}
{Matteucci}, F. 2021, \aapr, 29, 5

\bibitem[{{McConnachie} {et~al.}(2018){McConnachie}, {Ibata}, {Martin},
  {Ferguson}, {Collins}, {Gwyn}, {Irwin}, {Lewis}, {Mackey}, {Davidge},
  {Arias}, {Conn}, {C{\^o}t{\'e}}, {Crnojevic}, {Huxor}, {Penarrubia},
  {Spengler}, {Tanvir}, {Valls-Gabaud}, {Babul}, {Barmby}, {Bate}, {Bernard},
  {Chapman}, {Dotter}, {Harris}, {McMonigal}, {Navarro}, {Puzia}, {Rich},
  {Thomas}, \& {Widrow}}]{mcc18}
{McConnachie}, A.~W., {Ibata}, R., {Martin}, N., {et~al.} 2018, \apj, 868, 55

\bibitem[{{McConnachie} {et~al.}(2009){McConnachie}, {Irwin}, {Ibata},
  {Dubinski}, {Widrow}, {Martin}, {C{\^o}t{\'e}}, {Dotter}, {Navarro},
  {Ferguson}, {Puzia}, {Lewis}, {Babul}, {Barmby}, {Bienaym{\'e}}, {Chapman},
  {Cockcroft}, {Collins}, {Fardal}, {Harris}, {Huxor}, {Mackey},
  {Pe{\~n}arrubia}, {Rich}, {Richer}, {Siebert}, {Tanvir}, {Valls-Gabaud}, \&
  {Venn}}]{mcc09}
{McConnachie}, A.~W., {Irwin}, M.~J., {Ibata}, R.~A., {et~al.} 2009, \nat, 461,
  66

\bibitem[{{Monachesi} {et~al.}(2019){Monachesi}, {G{\'o}mez}, {Grand},
  {Simpson}, {Kauffmann}, {Bustamante}, {Marinacci}, {Pakmor}, {Springel},
  {Frenk}, {White}, \& {Tissera}}]{Monachesi2019}
{Monachesi}, A., {G{\'o}mez}, F.~A., {Grand}, R. J.~J., {et~al.} 2019, \mnras,
  485, 2589

\bibitem[{{Nomoto} {et~al.}(2013){Nomoto}, {Kobayashi}, \&
  {Tominaga}}]{Nomoto13}
{Nomoto}, K., {Kobayashi}, C., \& {Tominaga}, N. 2013, \araa, 51, 457

\bibitem[{Oliphant(2015)}]{numpy}
Oliphant, T.~E. 2015, Guide to NumPy, 2nd edn. (USA: CreateSpace Independent
  Publishing Platform)

\bibitem[{{Pe{\~n}a} \& {Flores-Dur{\'a}n}(2019)}]{Pena19}
{Pe{\~n}a}, M. \& {Flores-Dur{\'a}n}, S.~N. 2019, \rmxaa, 55, 255

\bibitem[{{Pulsoni} {et~al.}(2018){Pulsoni}, {Gerhard}, {Arnaboldi}, {Coccato},
  {Longobardi}, {Napolitano}, {Moylan}, {Narayan}, {Gupta}, {Burkert},
  {Capaccioli}, {Chies-Santos}, {Cortesi}, {Freeman}, {Kuijken}, {Merrifield},
  {Romanowsky}, \& {Tortora}}]{pul18}
{Pulsoni}, C., {Gerhard}, O., {Arnaboldi}, M., {et~al.} 2018, \aap, 618, A94

\bibitem[{{Quinn} \& {Goodman}(1986)}]{Quinn86}
{Quinn}, P.~J. \& {Goodman}, J. 1986, \apj, 309, 472

\bibitem[{{Sanders} {et~al.}(2012){Sanders}, {Caldwell}, {McDowell}, \&
  {Harding}}]{san12}
{Sanders}, N.~E., {Caldwell}, N., {McDowell}, J., \& {Harding}, P. 2012, \apj,
  758, 133

\bibitem[{Schwarz(1978)}]{Schwarz78}
Schwarz, G. 1978, The Annals of Statistics, 6, 461

\bibitem[{{Scott} {et~al.}(2021){Scott}, {van de Sande}, {Sharma},
  {Bland-Hawthorn}, {Freeman}, {Gerhard}, {Hayden}, \& {McDermid}}]{Scott21}
{Scott}, N., {van de Sande}, J., {Sharma}, S., {et~al.} 2021, \apjl, 913, L11

\bibitem[{{Sellwood}(2014)}]{Sellwood14}
{Sellwood}, J.~A. 2014, Reviews of Modern Physics, 86, 1

\bibitem[{{Spitoni} {et~al.}(2019){Spitoni}, {Silva Aguirre}, {Matteucci},
  {Calura}, \& {Grisoni}}]{Spitoni19}
{Spitoni}, E., {Silva Aguirre}, V., {Matteucci}, F., {Calura}, F., \&
  {Grisoni}, V. 2019, \aap, 623, A60

\bibitem[{{Spitoni} {et~al.}(2021){Spitoni}, {Verma}, {Silva Aguirre},
  {Vincenzo}, {Matteucci}, {Vai{\v{c}}ekauskait{\.{e}}}, {Palla}, {Grisoni}, \&
  {Calura}}]{Spitoni21}
{Spitoni}, E., {Verma}, K., {Silva Aguirre}, V., {et~al.} 2021, \aap, 647, A73

\bibitem[{{Stanghellini} \& {Haywood}(2018)}]{Stanghellini18}
{Stanghellini}, L. \& {Haywood}, M. 2018, \apj, 862, 45

\bibitem[{{Ventura} {et~al.}(2017){Ventura}, {Stanghellini}, {Dell'Agli}, \&
  {Garc{\'\i}a-Hern{\'a}ndez}}]{Ventura17}
{Ventura}, P., {Stanghellini}, L., {Dell'Agli}, F., \&
  {Garc{\'\i}a-Hern{\'a}ndez}, D.~A. 2017, \mnras, 471, 4648

\bibitem[{{Virtanen} {et~al.}(2019){Virtanen}, {Gommers}, {Oliphant},
  {Haberland}, {Reddy}, {Cournapeau}, {Burovski}, {Peterson}, {Weckesser},
  {Bright}, {van der Walt}, {Brett}, {Wilson}, {Jarrod Millman}, {Mayorov},
  {Nelson}, {Jones}, {Kern}, {Larson}, {Carey}, {Polat}, {Feng}, {Moore}, {Vand
  erPlas}, {Laxalde}, {Perktold}, {Cimrman}, {Henriksen}, {Quintero}, {Harris},
  {Archibald}, {Ribeiro}, {Pedregosa}, {van Mulbregt}, \&
  {Contributors}}]{scipy}
{Virtanen}, P., {Gommers}, R., {Oliphant}, T.~E., {et~al.} 2019, arXiv
  e-prints, arXiv:1907.10121

\bibitem[{{Westmeier} {et~al.}(2008){Westmeier}, {Br{\"u}ns}, \&
  {Kerp}}]{Westmeier08}
{Westmeier}, T., {Br{\"u}ns}, C., \& {Kerp}, J. 2008, \mnras, 390, 1691

\bibitem[{{Williams} {et~al.}(2017){Williams}, {Dolphin}, {Dalcanton}, {Weisz},
  {Bell}, {Lewis}, {Rosenfield}, {Choi}, {Skillman}, \& {Monachesi}}]{wil17}
{Williams}, B.~F., {Dolphin}, A.~E., {Dalcanton}, J.~J., {et~al.} 2017, \apj,
  846, 145

\bibitem[{{Wisnioski} {et~al.}(2015){Wisnioski}, {F{\"o}rster Schreiber},
  {Wuyts}, {Wuyts}, {Bandara}, {Wilman}, {Genzel}, {Bender}, {Davies},
  {Fossati}, {Lang}, {Mendel}, {Beifiori}, {Brammer}, {Chan}, {Fabricius},
  {Fudamoto}, {Kulkarni}, {Kurk}, {Lutz}, {Nelson}, {Momcheva}, {Rosario},
  {Saglia}, {Seitz}, {Tacconi}, \& {van Dokkum}}]{Wisnioski15}
{Wisnioski}, E., {F{\"o}rster Schreiber}, N.~M., {Wuyts}, S., {et~al.} 2015,
  \apj, 799, 209

\bibitem[{{Yoachim} \& {Dalcanton}(2006)}]{Yoachim06}
{Yoachim}, P. \& {Dalcanton}, J.~J. 2006, \aj, 131, 226

\bibitem[{{Yoachim} \& {Dalcanton}(2008)}]{Yoachim08}
{Yoachim}, P. \& {Dalcanton}, J.~J. 2008, \apj, 683, 707

\bibitem[{{Zahid} {et~al.}(2017){Zahid}, {Kudritzki}, {Conroy}, {Andrews}, \&
  {Ho}}]{Zahid17}
{Zahid}, H.~J., {Kudritzki}, R.-P., {Conroy}, C., {Andrews}, B., \& {Ho}, I.~T.
  2017, \apj, 847, 18

\bibitem[{{Zurita} \& {Bresolin}(2012)}]{Zurita12}
{Zurita}, A. \& {Bresolin}, F. 2012, \mnras, 427, 1463

\end{thebibliography}

\begin{appendix} 

\section{AGB evolution and possible dependencies of PN oxygen and argon abundances on progenitor mass and metallicity}
\label{sect:agb}

\begin{figure}
        \centering
        \includegraphics[width=\columnwidth,angle=0]{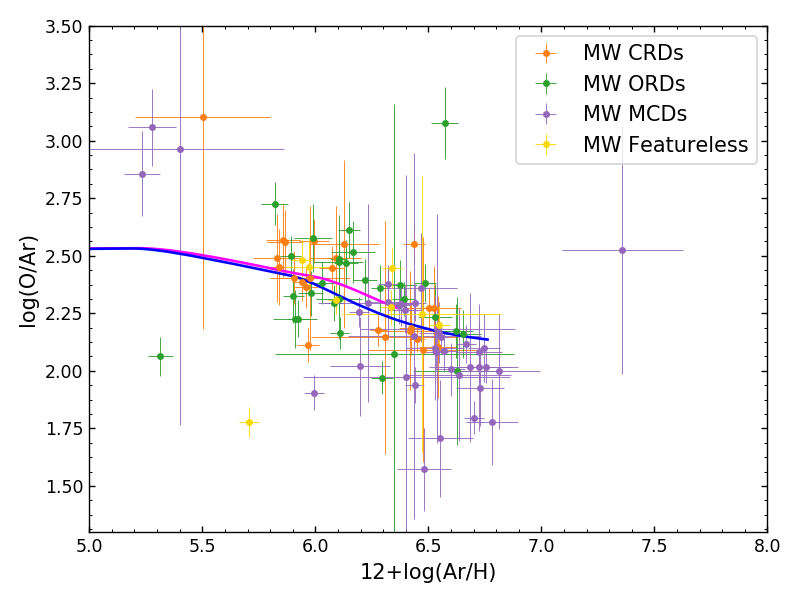}
        \caption{log(O/Ar) vs. 12 + log(Ar/H) distribution of 101 MW PNe marked by their circumstellar dust types from \citet{Ventura17}. The chemical evolution model tracks for the MW are same as in Figure~\ref{fig:chem}.}
        \label{fig:sample2_dust}
\end{figure}

Measured argon abundances in PNe reflect the ISM abundance at the time of their birth as surface argon is invariant during AGB evolution \citep{Delgado-Inglada15,Garcia-Hernandez16}. However, surface oxygen has been found to be modified in the AGB phase, particularly in stars with initial mass $\geq3\rm M_{\odot}$ where hot-bottom burning (HBB) may result in an oxygen depletion of up to $\sim0.2$ dex \citep[e.g.][]{Garcia-Hernandez16, Ventura17}. This results in measured PN oxygen abundances lower than their birth values. On the other hand, for PNe evolving from stars with initial masses of $1-2\rm M_{\odot}$ and Z$<0.008$, third dredge-up (TDU) effects may result in an oxygen enrichment of up to $\sim0.3$ dex \citep[e.g.][]{Garcia-Hernandez16, Ventura17}. 

In a small sample of 20 MW PNe, \citet[][]{Delgado-Inglada15} had found that oxygen is enriched by up to $\sim0.3$ dex for intermediate metallicities of 12+(O/H) = 8.2--8.7 for PNe with Carbon-rich (circumstellar) dust (CRDs), while it is invariant in PNe with oxygen-rich (circumstellar) dust (ORDs). In \citetalias{Bhattacharya22} (see their Appendix D for a detailed discussion), using a larger sample of 101 MW PNe with abundance measurements and dust characterisation compiled by \citet{Ventura17} compared to the chemical evolution tracks by \citet{Kobayashi20}, we found that both CRDs and ORDs (as well as PNe with featureless dust) follow the MW stellar evolution tracks with no oxygen enrichment or depletion relative to argon abundances. The same can be seen in Figure~\ref{fig:sample2_dust}. However, as also noted \citetalias{Bhattacharya22}, many of the MW PNe with
mixed chemistry dust (MCDs) that are metal-rich (12 + log(Ar/H)
> 6.3) preferentially have log(O/Ar) values below the model tracks,
indicating lower oxygen or oxygen depletion. These MCDs are the youngest (<300 Myr) most metal-rich PNe in the MW sample \citep{GarciaHernandez14}.

\begin{figure}
        \centering
        \includegraphics[width=\columnwidth,angle=0]{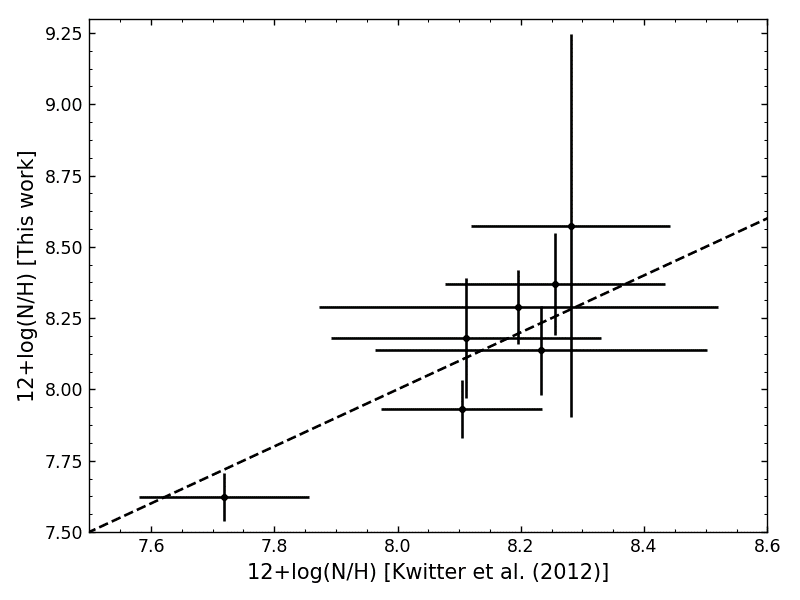}
        \includegraphics[width=\columnwidth,angle=0]{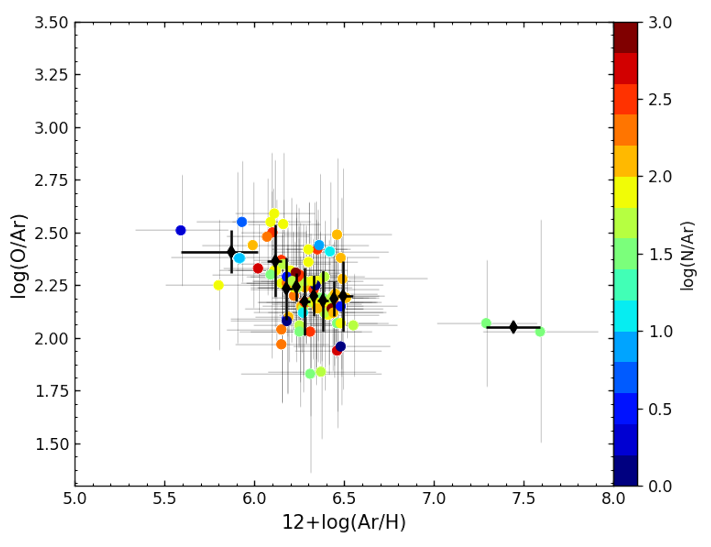}
        \caption{[Top] Nitrogen abundance measurements compared for 7 M~31 disc PNe between \citet{Kwitter21} and this work. [Bottom] log(O/Ar) vs. 12 + log(Ar/H) distribution of 91 M~31 disc PNe coloured by their log(N/Ar) values. Measurement uncertainties are shown in grey.The PNe are binned as a function of their 12+log(Ar/H) values. For each bin the mean log(O/Ar) is plotted as black diamonds while the error bar shows their standard deviation.}
        \label{fig:nitro}
\end{figure}

As such young PNe are expected to exhibit enhanced nitrogen abundances based on the AGB evolution models of their progenitors \citep{GarciaHernandez14,Ventura17}, we can check whether any M~31 PN has enhanced nitrogen as further indication of a very young age and oxygen depletion, if any.

We measure nitrogen ionic abundances from the [\ion{N}{II}] 6548,6584~\AA~ line fluxes for a sub-sample of 87 M~31 disc PNe which were observed in 2018 and 2019 with Hectospec at the MMT. The nitrogen elemental abundance is obtained by using the ICF from \citet{Delgado14} (the nitrogen abundances for the full sample will be presented in a future publication). The nitrogen abundances for 7 PNe in the M~31 disc are also measured by \citet{Kw12} using Cloudy photoionisation models, independent of ICF. Figure~\ref{fig:nitro} [Top] shows that the nitrogen abundance measurements of these 7 PNe derived here are consistent with those of \citet{Kw12}, indicating that the Nitrogen abundance measurements are not dependent strongly on the adopted ICF. 

Figure~\ref{fig:nitro} [Bottom] shows the log(O/Ar) vs. 12 + log(Ar/H) distribution of these PNe coloured by their nitrogen-to-argon abundance ratio, log(N/Ar). Young PNe, possibly affected by HBB, would show higher values of log(N/Ar) than the general trend, owing to their expected enhanced nitrogen abundances as well as low log(O/Ar) values due to oxygen depletion. To obtain the general trend of log(O/Ar) values as a function of 12+log(Ar/H), as in Section~\ref{sect:trends}, we divide the PNe in bins of 12+log(Ar/H) such that there are $10$ PNe in each bin (the bins with the smallest and largest argon abundance have fewer measurements, five and two respectively). We then calculate the mean and standard deviation of the log(O/Ar) values of PNe in each 12+log(Ar/H) bin, shown as diamonds in Figure~\ref{fig:nitro} [Bottom]. The PNe with the highest 25-percentile log(N/Ar) values have a mean offset of 0.02~dex ($\sigma\rm_{offset}=0.11$~dex) from the general trend, while those with the lowest 75-percentile log(N/Ar) values have a mean offset of 0.01~dex ($\sigma\rm_{offset}=0.14$~dex). Therefore, the PNe with higher log(N/Ar) values do not preferentially occupy lower log(O/Ar) values below the general trend. We can thus state that the number of PNe affected by HBB in our M~31 subsample is very small and does not drive the measured log(O/Ar) trends. This is consistent with the M~31 low- and high- extinction PNe having average ages $\sim4.5$~Gyr and $\sim2.5$~Gyr respectively with the bulk of the latter having likely formed in a burst of star formation $\sim$2 Gyr ago (\citetalias{Bh+19b}), implying therefore a very small number of PNe with very young massive progenitors (affected by HBB).

To summarise, we find no conclusive evidence of AGB evolution effects with modification of the oxygen abundance in the nebula to be driving the trends in log(O/Ar) vs. 12 + log(Ar/H) for the M~31 disc PNe studied in this work. Any such effect is within the measurement errors. We thus conclude the oxygen abundance measurements for M~31 PNe reflect their birth ISM chemical abundances, within the errors.

\section{Chemical evolution models}
\label{sect:chem-evol-W17}

\begin{figure*}
        \centering
        \includegraphics[width=1.8\columnwidth,angle=0]{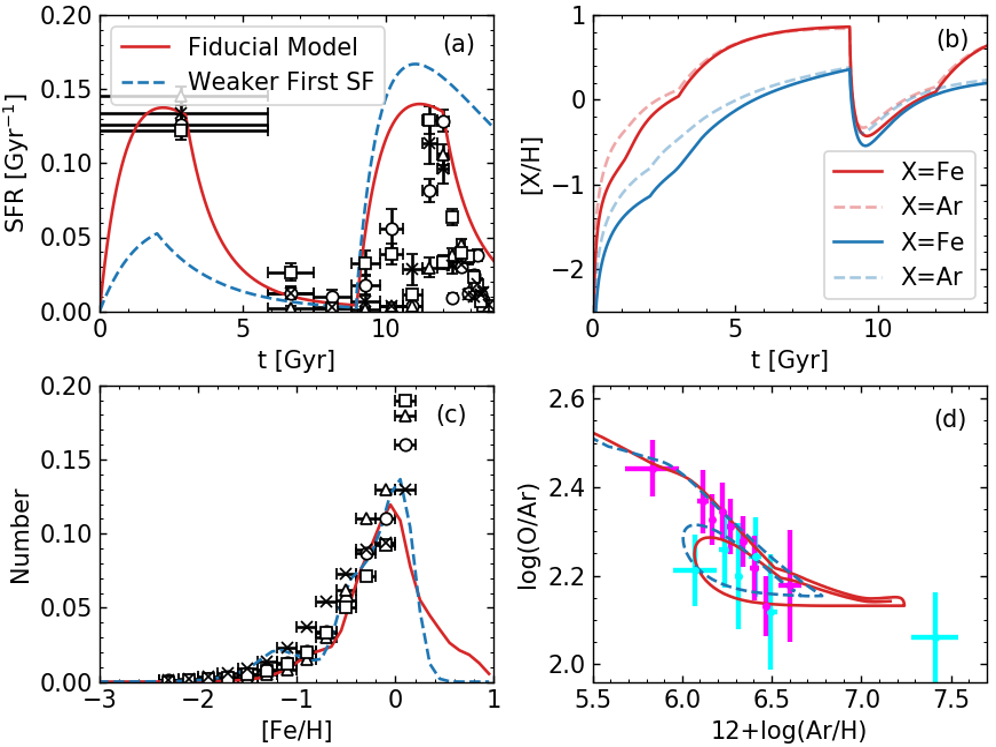}
        \caption{(a) Star formation history, comparing to the observational results from \citet{wil17}. (b) Evolution of iron (red and blue lines) and argon (light red and blue lines) abundance in the ISM. (c) Metallicity distribution function, comparing to the observational results from \citet{wil17}. (d) The O/Ar--Ar/H relation, comparing to our PN sample; magenta for low-extinction PNe over the entire M31 disc ($R_{GC}=$2--30 kpc) and cyan for high-extinction PNe at $R_{GC}<14$ kpc {(same as in Figure~\ref{fig:loop_model})}.}
        \label{fig:chem-last}
\end{figure*}

\begin{table}
\caption{Timescales of the two infall models for the M31 $2 \le R_{\rm GC} \le 14$ kpc region.} 
\centering
\adjustbox{max width=\columnwidth}{
\begin{tabular}{l|ccccccc}\hline
& $t_1$ & $t_2$ & $t_3$ & $\tau_{\rm i,1}=\tau_{\rm i,2}$ & $\tau_{\rm s,1}$ & $\tau_{\rm s,2}$ & $\tau_{\rm o,1}=\tau_{\rm o,2}$\\ \hline
fiducial model & 3 & 9 & 12 & 5 & 1 & 1 & 4 \\
weaker 1st SF  & 2 & 9 & -  & 5 & 5 & 1 & 3 \\
outer disc & - & - & - & 1 & 1 & - & - \\ \hline
\end{tabular}
}
\label{tab:chem_mod}
\tablefoot{{  This table summarised the timescales of the two infall models. The units are in Gyr. The timescales are: the timescales of the infall ($\tau_{\rm i}$), that of star formation ($\tau_{\rm s}$), and outflow (if there is, $\tau_{\rm o}$). The table lists three epochs that identify the truncation of the first infall (at $t_1$), the onset of the second infall (at $t_2$), and the truncation of the second infall (at $t_3$), also. See Appendix~\ref{sect:chem-evol-W17} for more details. }
}
\end{table}

Previous works to model thin and thick discs used the framework of classical, one-zone galactic chemical evolution (GCE) models. \citet{Grisoni17} favoured a parallel model in which thin and thick discs form simultaneously. The thin and thick disc models of the MW in \citet{Kobayashi20}, which are plotted in Figures 3 and 4, also follow this approach. The thick disc stars form with shorter star formation and chemical enrichment timescales than for the thin disc stars, and hence in the [$\alpha$/Fe]--[Fe/H] or O/Ar--Ar/H diagrams, the thick disc model appears above the thin disc model. There is a small number of old metal-poor stars in the thin disc model. On the other hand, following \citet{Chiappini97}, \citet{Spitoni19} concluded that a two-infall model in which the thick disc form before the thin disc, with a long delay until the second infall, is better. Our models for M31 inner disc plotted in Figure~\ref{fig:loop_model} also assumes two infalls. 

In our two-infall models, it is assumed that two exponential infalls of primordial gas trigger two star bursts. The star formation rate is assumed to be proportional to the gas fraction (see \citealt{Kobayashi20} for the formula).
Timescales of the infall ($\tau_{\rm i}$), star formation ($\tau_{\rm s}$), and outflow (if there is, $\tau_{\rm o}$) are determined separately. In addition three epochs are set for the truncation of the first infall (at $t_1$), the onset of the second infall (at $t_2$), and the truncation of the second infall (at $t_3$). {  The IMF slope is assumed to be invariant (Kroupa IMF with a slope $x=1.3$ at the massive end), so as for calculating the nucleosynthesis yields (\citet{Kobayashi20} for single stars including AGB stars and core-collapse supernovae, and \citet{Kobayashi20b} for Type Ia supernovae) and Type Ia supernova progenitor model \citep{Kobayashi09}.}
Therefore, there are nine free parameters at maximum. However, it is possible to choose the best set of parameters if the following observational constrains are available: the metallicity distribution function (MDF), star formation rates (i.e., ages of stars), and elemental abundance ratios. 

We take the observational results from \citet{wil17}, whose sample covers the M31 inner disc and we combine them to our O/Ar--Ar/H relation at {   $2 \le R_{\rm GC} \le 14$ kpc.} Calculating about 100 GCE models, we found this fiducial model (the solid line in Figure~\ref{fig:chem-last}), which shows two similar star bursts with the same height (panel a). Note that the observational results highly depends on the input stellar evolution models (open circles for Padova, triangles for BaSTI, squares for PARSEC, and crosses for MIST).

The first star burst increases the metallicity quickly, which exceeds the super-solar metallicity at $t \sim 3$ Gyr. Then the metallicity suddenly decrease by the second infall at $t=9$ Gyr, of which chemical composition is assumed to be primordial. Needless to say, with a metal-enhanced infall, the metallicity drop becomes smaller. During the break of infall ($t=3$ to $9$ Gyr), star formation is much suppressed, but the iron abundance keeps increasing (Figure~\ref{fig:chem-last}, panel b) due to the delayed enrichment from Type Ia supernovae, which is one of the two important factors to avoid the overproduction of metal-poor stars in the MDF (the G-dwarf problem, panel c). The other factor is a long infall timescale. Moderate outflow is also included, proportional to the star formation rate, in order to prevent the metallicity to become too high. The adopted parameters are summarized in Table~\ref{tab:chem_mod}. As discussed, since Ar is also produced by Type Ia supernovae \citep{Kobayashi20b}, the [Ar/H] evolution follows well the [Fe/H] evolution (panel b).

The O/Ar--Ar/H relation is shown in Figure~\ref{fig:chem-last} panel d. The plateau value is determined from the IMF-weighted yields of core-collapse supernovae. The decreasing trend is caused by 1) the knee at $12 + \log(Ar/H)  = 5.3$ caused by the dependency of argon production on the metallicity of core collapse Type II SNe, and 2) the delayed enrichment from Type Ia supernovae at $12 + \log(Ar/H)  = 6.0$ . A loop is created by the second infall, followed by the second star burst. With outflow, the loop becomes larger. The blue dashed line is an alternative model with a weaker first star formation. Although this model produce fewer old stars in Figure~\ref{fig:chem-last} panel a, it gives a better match in Figure~\ref{fig:chem-last} panel c and d. Namely the lack of metal-rich low-O/Ar PNe prefers a weaker first star burst.

{  In our thin disc sample at $R_{\rm GC}>18$ kpc,} the O/Ar seem to be higher than these model predictions. This can be explained if the outer {thin} disc formed with a shorter star formation timescale, as in the model plotted in Figure~\ref{fig:outerdisc}. The adopted timescale is even shorter than the MW thick disc and is as short as for nearby early-type galaxies.

\end{appendix}
\end{document}